\documentclass{sigchi}
\usepackage{array}
\usepackage{threeparttable}
\usepackage{enumitem}
\usepackage{caption}
\toappear{\scriptsize Permission to make digital or hard copies of all or part of this work for personal or classroom use is granted without fee provided that copies are not made or distributed for profit or commercial advantage and that copies bear this notice and the full citation on the first page. Copyrights for components of this work owned by others than the author(s) must be honored. Abstracting with credit is permitted. To copy otherwise, or republish, to post on servers or to redistribute to lists, requires prior specific permission and/or a fee. Request permissions from permissions@acm.org. \\
{\emph{CHI '20, April 25--30, 2020, Honolulu, HI, USA.} } \\
Copyright is held by the owner/author(s). Publication rights licensed to ACM. \\
ACM ISBN 978-1-4503-6708-0/20/04\ ...\$15.00.\\
http://dx.doi.org/10.1145/3313831.3376229}
\usepackage{balance}       % to better equalize the last page
\usepackage{graphics}      % for EPS, load graphicx instead 
\usepackage[T1]{fontenc}   % for umlauts and other diaeresis
\usepackage{txfonts}
\usepackage{mathptmx}
\usepackage[pdflang={en-US},pdftex]{hyperref}
\usepackage{color}
\usepackage{booktabs}
\usepackage{textcomp}

\usepackage{microtype}        % Improved Tracking and Kerning
\usepackage{ccicons}          % Cite your images correctly!
\usepackage{todonotes}

\def\plaintitle{A Human-Centered Review of the Algorithms used within the U.S. Child Welfare System}

\def\emptyauthor{}
\def\plainkeywords{Child Welfare System; Algorithmic Decision-Making; Human-centered Algorithm Design}

\makeatletter
\def\url@leostyle{%
  \@ifundefined{selectfont}{
    \def\UrlFont{\sf}
  }{
    \def\UrlFont{\small\bf\ttfamily}
  }}
\makeatother
\def\pprw{8.5in}
\def\pprh{11in}

\definecolor{linkColor}{RGB}{6,125,233}
\hypersetup{%
  pdftitle={\plaintitle},
  pdfauthor={\emptyauthor},
  pdfkeywords={\plainkeywords},
  pdfdisplaydoctitle=true, % For Accessibility
  bookmarksnumbered,
  pdfstartview={FitH},
  colorlinks,
  citecolor=black,
  filecolor=black,
  linkcolor=black,
  urlcolor=linkColor,
  breaklinks=true,
  hypertexnames=false
}

\begin{document}
\title{\plaintitle}

\numberofauthors{4}
\author{%
  \alignauthor{Devansh Saxena\\
    \affaddr{Marquette University}\\
    \affaddr{Milwaukee, WI, USA}\\
    \email{devansh.saxena@marquette.edu}}\\
  \alignauthor{Karla Badillo-Urquiola\\
    \affaddr{University of Central Florida}\\
    \affaddr{Orlando, FL, USA}\\
    \email{kbadillo@ist.ucf.edu}}\\
  \alignauthor{Pamela J. Wisniewski\\
    \affaddr{University of Central Florida}\\
    \affaddr{Orlando, FL, USA}\\
    \email{pamwis@ucf.edu}}\\
  \alignauthor{Shion Guha\\
    \affaddr{Marquette University}\\
    \affaddr{Milwaukee, WI, USA}\\
    \email{shion.guha@marquette.edu}}
%   \alignauthor{Leave Authors Anonymous\\
%     \affaddr{for Submission}\\
%     \affaddr{City, Country}\\
%     \email{e-mail address}}\\
%   \alignauthor{Leave Authors Anonymous\\
%     \affaddr{for Submission}\\
%     \affaddr{City, Country}\\
%     \email{e-mail address}}\\
%   \alignauthor{Leave Authors Anonymous\\
%     \affaddr{for Submission}\\
%     \affaddr{City, Country}\\
%     \email{e-mail address}}\\
}

\maketitle

\begin{abstract}
The U.S. Child Welfare System (CWS) is charged with improving outcomes for foster youth; yet, they are overburdened and underfunded. To overcome this limitation, several states have turned towards algorithmic decision-making systems to reduce costs and determine better processes for improving CWS outcomes. Using a human-centered algorithmic design approach, we synthesize 50 peer-reviewed publications on computational systems used in CWS to assess how they were being developed, common characteristics of predictors used, as well as the target outcomes. We found that most of the literature has focused on risk assessment models but does not consider theoretical approaches (e.g., child-foster parent matching) nor the perspectives of caseworkers (e.g., case notes). Therefore, future algorithms should strive to be context-aware and theoretically robust by incorporating salient factors identified by past research. We provide the HCI community with research avenues for developing human-centered algorithms that redirect attention towards more equitable outcomes for CWS.
\end{abstract}

% ACM Classfication

\begin{CCSXML}
<ccs2012>
<concept>
<concept_id>10003120.10003121</concept_id>
<concept_desc>Human-centered computing~Human computer interaction (HCI)</concept_desc>
<concept_significance>500</concept_significance>
</concept>
<concept>
<concept_id>10003120.10003121.10003125.10011752</concept_id>
<concept_desc>Human-centered computing~Haptic devices</concept_desc>
<concept_significance>300</concept_significance>
</concept>
<concept>
<concept_id>10003120.10003121.10003122.10003334</concept_id>
<concept_desc>Human-centered computing~User studies</concept_desc>
<concept_significance>100</concept_significance>
</concept>
</ccs2012>
\end{CCSXML}

\ccsdesc[500]{Applied computing~Computing in government}
\ccsdesc[500]{Information systems~Decision support systems}

% Author Keywords
\keywords{\plainkeywords}

% Print the classficiation codes
\printccsdesc
%Please use the 2012 Classifiers and see this link to embed them in the text: \url{https://dl.acm.org/ccs/ccs_flat.cfm}

\section{Introduction}
As of September 2016, there were 437,465 children in the child welfare system (CWS) in United States \cite{us2017afcars}. This is a significant (10\%) rise in just 4 years since September 2012 \cite{us2017afcars}, and this number is expected to keep rising unless significant efforts are made to improve youth outcomes \cite{us2017afcars}. Child abuse and neglect are severe issues that policymakers in the United States continue to battle with, and which is consistently at the foreground of public policy \cite{congress2008fostering}. In recent years, CWS has  been the center of public and media scrutiny \cite{cooper2005implications} because of the potential damage done to the children who are removed from the care of their parents \cite{davoren1975foster}. Therefore, there is significant pressure on CWS to systematize the decision-making process and show that these decisions were unbiased and evidenced-based \cite{noonan2009legal}. For most policymakers, algorithmic decisions are perceived to be the epitome of being unbiased, evidence-based, and objective \cite{tushnet2018difference, congressbill}. Thus, algorithms have been developed for almost every aspect of services provided by CWS in different states. For instance, models have been developed for predicting risk of future maltreatment event of a child \cite{vaithianathan2017developing}, recommending appropriate placement settings \cite{schwab1984matching} and matching children with foster parents who can meet the unique needs of every child \cite{moore2016assessing}. Many of these algorithms have achieved various degrees of early successes and have shown to reduce costs \cite{ringel2018improving} for CWS. However, they have also come under significant criticisms for being biased \cite{church2017search, berger2009estimating}, being opaque \cite{tushnet2018difference}, complex and hard to explain \cite{vaithianathan2017developing, chouldechova2018case}, being too reductive \cite{cohen2014applied} and non-contextual \cite{schwalbe2008strengthening} and for not incorporating factors that arise from relevant social science research literature \cite{carnochan2013achieving}. 

The SIGCHI community is at the forefront of research on algorithmic bias \cite{danks2017algorithmic, bozdag2013bias, lambrecht2019algorithmic}, and has begun to examine some of the challenges of algorithmic decision-making within CWS. Brown et al. \cite{brown2019toward} studied community perspectives on algorithmic decision-making systems in CWS and found several aspects of algorithmic systems that bolstered distrust, perpetuated bias, concern over the lack of contextual understanding and `black-box' nature of the algorithms, as well as concerns about how these algorithms may negatively impact child-welfare workers' decisions. Moreover, scholars outside of HCI have discussed how algorithms impact decision-making in CWS \cite{camasso2013decision, schwalbe2008strengthening, shlonsky2005next, fowler2019scaling}. Engaging in research that helps people and organizations, such as CWS, is well-suited and important for the HCI community. Therefore, a critical step in building a strategic research agenda is to synthesize the breadth of work that has already been done to identify a pathway forward. To forge this path, we posed the following high-level research questions:

\noindent \textbf{RQ1:} \textit{What methods have been used to build algorithms in the child welfare system?}

\vspace{-2mm}
\noindent \textbf{RQ2:} \textit{What factors (i.e., independent variables) have been shown to be salient in predicting CWS outcomes?}

\vspace{-2mm}
\noindent \textbf{RQ3:} \textit{What outcomes (i.e., dependent variables) have CWS organizations been predicting?}

To answer these questions, we conducted a comprehensive literature review (n=50) of algorithms used for decision-making in CWS in the United States. We qualitatively analyzed these articles using the lens of human-centered algorithm design \cite{baumer2017toward}. Overall, we found that majority of the algorithms in CWS are empirically constructed, even though the empirical knowledge is quite fragmented \cite{gambrill2001need}. Our results also revealed considerable differences in the predictors currently being used and those found salient in the child-welfare literature. Finally, CWS has  traditionally focused on `risk assessment,' rather than positive outcomes that improve the lives of foster children. Based on Woobrock and Kientz's encapsulation of research contributions in HCI \cite{wobbrock2016research}, this paper is a survey of the existing literature and makes the following unique research contributions:
\begin{enumerate}
    \itemsep-0em
    \item We apply a human-centered conceptual framework \cite{baumer2017toward} to critically review the algorithms used within the U.S. child welfare system.
    \item We introduce domain knowledge from the child welfare system to embed it within the SIGCHI community to allow for collaborative research between the two disciplines.
    \item We identify the potential gaps in the existing literature and recommend future research opportunities with careful attention to the human-centered design of algorithms to benefit CWS.
\end{enumerate}

In the following sections, we discuss Human-Centered Algorithm Design and how we used this framework to inform our literature review methodology. Next, we situate our research within the SIGCHI community.

\vspace{-0.5em}
\subsection{A Human-Centered Approach to Algorithm Design}
As algorithms begin to permeate through every aspect of social life, HCI researchers have begun to ask, "Where is the Human?", that is, recognizing that humans are a critical, if not the central component of many domains for which Artificial Intelligence (AI) systems are being developed. A workshop organized at CHI 2019 \cite{inkpen2019human}, tackled this topic to identify several pertinent issues in algorithmic design, such as the opaque and isolated development of algorithms and a lack of involvement of the human stakeholders, who use these systems and are most affected by them. To address these problems, Baumer proposed Human-Centered Algorithm Design (HCAD) \cite{baumer2017toward}; a conceptual framework founded in practices derived from human-centered design \cite{luma2012innovating}. It incorporates human and social interpretations through both the design and evaluation phases \cite{baumer2017toward}. Baumer \cite{baumer2017toward} lays out three strategies that help algorithm design become more human-centered, namely, 1) theoretical, 2) speculative and 3) participatory strategies. We draw from the theoretical perspective to frame our research questions and as the qualitative lens for our analysis. Human-centered theoretical design strategy informs algorithm design as follows:
\begin{itemize}
    \setlength\itemsep{0em}
    \item Meaning-making: Theoretical foundations provide a much-needed scaffolding for dealing with complexity, identifying and evaluating design opportunities \cite{pink2013applying}. Designers need to study the socio-cultural domain in which they intend to situate their work.    
    \item Design:  Theoretical approaches aim to incorporate concepts and theories from social sciences into data science \cite{baumer2017toward}. 
    \item Evaluation: The stakeholders' social interpretations of results can help ensure that the algorithm has higher utility and integrates well with practice. 
\end{itemize}
CWS is one such domain that suffers from a complete lack of human perspectives through the design process. Therefore, our work focuses on how HCAD strategies can be employed to answer critical research questions in CWS.

% \begin{itemize}
%     \setlength\itemsep{0em}
%     \item Theoretical strategy - Theoretical approaches can be used to incorporate pertinent factors arising from social science literature. Theory can be used to not only inform feature selection (design) but also interpret the results (evaluation).
%     \item Speculative strategy - This design strategy allows researchers to think outside the box such that they are not restrained by what is technically feasible. The idea is to provoke thought and dialog rather than produced finished products. This is especially useful in case of algorithms where the range of possibilities changes quite rapidly.
%     \item Participatory strategy - This strategy advocates for the active involvement of stakeholders through both the design and evaluation processes. Addressing the concerns of stakeholders through the design process leads to AI systems that are both usable and useful \cite{luma2012innovating}.
% \end{itemize}

\section{Background}
We situate our research within the SIGCHI community and provide an overview of the work that has been done to develop integrated data systems for CWS.

\vspace{-0.5em}
\subsection{SIGCHI Research to Support the Child Welfare System}
The SIGCHI community has recognized the importance of conducting research with organizations that help disadvantaged communities, such as those experiencing homelessness \cite{strohmayer2015exploring, woelfer2010homeless} or recovering from substance abuse \cite{maclean2015forum77}. For example, Strohmayer, Comber, and Balaam \cite{strohmayer2015exploring} partnered with a center for people of low social stability to understand homeless young adults' perceptions of education. Similarly, Woelfer and Hendry \cite{woelfer2010homeless} created a community technology center at a local service agency to work with homeless young people, case managers, and outreach workers. Similarly, SIGCHI researchers have started to engage with CWS to find ways to improve the lives of youth who have been displaced from their families. 
Some SIGCHI research has focused on foster youth and parents. For instance, Gray et al.'s \cite{gray2019trove} research with fostered and adopted children introduces a new digital memory box for creating and storing childhood memories. More recently, researchers have begun to study algorithmic decision-making systems within the child-welfare community. Badillo-Urquiola et al. \cite{badillo2018chibest} presented the challenges foster parents face mediating teens' technology use within the home. 

Most relevant to our current work, Brown et al. \cite{brown2019toward} engaged in a participatory design effort and conducted workshops with families involved in CWS, child-welfare workers, and service providers. They found that participants were uncomfortable with algorithmic systems. Participants felt that these systems used deficit-based frameworks to make decisions and questioned the bias present within the data. Based on their findings, the investigators provide recommendations for researchers and designers to work together with public service agencies to develop systems that provide a higher comfort level to the community. Our study builds upon this related work by critically investigating the algorithms used within CWS and highlighting opportunities for future research. We provide a foundation for implementing human-centered approaches in the design and development of algorithmic systems for CWS. 
%Furthermore, there has also been research within the SIGCHI community that has studied technology within the context of the foster care system, but at the individual level. They note the dearth of research within this area and provide areas for further investigation. For example, research with fostered and adopted children introduces a new digital memory box for creating and storing childhood memories \cite{gray2019trove}. The researchers' proof-of-concept provide insights into the current challenges and frustrations of documenting and supporting a fostered or adopted child's life story. Other research conducted with foster parents has also presented the challenges foster parents face mediating their teens' technology use within the home \cite{badillo2018chibest}. Parents were desperate for resources on how to prevent their foster teens from engaging in high-risk online behaviors. These studies highlight the need for more research investigating technological interventions that can help support children within the foster care system as well as their parents. We add to this growing body of knowledge by engaging with prior work on computational methods within the foster care system to reveal future research opportunities and form a consortium of researchers who are interested in expanding this research domain. 

\vspace{-0.5em}
\subsection{Sociotechnical Systems for Child-Welfare}
In this section, we provide necessary background context about integrated data systems that laid the foundation for algorithmic work in CWS. In 1995, the federal government launched \textit{SACWIS} (State Automated Child Welfare Information System) initiative to provide states with a federally funded and automated case management tool. These data systems allow states to collect and maintain data for program management and informing their decision making \cite{kleinberg2015prediction}. States that implement \textit{SACWIS} must also report their data to federal databases, such as \textit{NCANDS} (National Child Abuse and Neglect Data System) \cite{acf2017} and AFCARS (Adoption and Foster Care Analysis and Reporting System) \cite{us2017afcars}, to allow for the continual curation of comprehensive national databases. These data systems became the foundation for actuarial risk assessment tools, which have been mandated into practice, even though controversy still remains as to whether these tools should override the judgment of case workers who are most knowledgeable about a particular child's case \cite{schwalbe2008strengthening, schwalbe2004re, shlonsky2005next, wagner1998using, camasso2013decision}. 

Past survey papers have analyzed algorithms in CWS from a macro perspective, focusing on their reliability and validity with respect to consensus-based or clinical risk assessment models \cite{schwalbe2008strengthening, camasso2013decision}. Yet, they do not examine the mathematical or human-centered construction of these algorithms, that is, the techniques, the variable sets, or the outcomes predicted. This is especially important in CWS because each case of child neglect or abuse is contextually different and cannot be evaluated using the same set of significant predictors derived empirically \cite{camasso2013decision}. To this end, we conducted a systematic literature review and identify the potential gaps in the literature with careful attention to the development of algorithms across time, as well as the methods and variable sets used. 

% We outline our specific research findings as follows--

% \begin{enumerate}
%     \itemsep0em
%     \item There exist inequalities within the literature in terms of particular computational methods used, predictors collected, and outcomes predicted.  
%     \item Most literature has focused on risk assessment models that fail to account for theoretical approaches found within social work literature.
%     \item There is a need for theoretically-driven models that are transparent, explainable, and account for the contextual judgments of caseworkers.
% \end{enumerate}

% In the following section, we describe the methodology we implemented to complete our systematic literature review.

\section{Methods}
We describe our scoping criteria, systematic literature search and data analysis process.

\vspace{-0.5em}
\subsection{Scoping Criteria: Defining Algorithms}
To understand how "algorithms" are used in CWS, we first need to contextualize what we mean by algorithms. We conceptualized "algorithms" through the lens of \textit{Street-level Algorithms}, a term recently coined by Alkhatib and Bernstein \cite{alkhatib2019street} in the HCI community. Street-level algorithms are algorithmically based systems that directly interact with and make on-the-ground decisions about human lives and welfare in a sociotechnical system \cite{alkhatib2019street}. From a more technical perspective, we use recent inclusive definitions \cite{donoho201750, kutz2013data} for a whole suite of computational methods from statistical modeling (for e.g., generalized linear models) and machine learning. This allowed us to take a holistic viewpoint toward most forms of quantitative data analysis in CWS. Statistical modeling and machine learning are not mutually exclusive but we differentiate between them based on assumptions made about the data as specified by Breiman \cite{breiman2001statistical}.

\vspace{-0.5em}
\subsection{Systematic Literature Search}
This study has been undertaken as a systematic literature review based on the guidelines proposed by Webster and Watson \cite{webster2002analyzing}. The unit of analysis for this literature review was peer-reviewed articles. We wanted to examine not just the algorithms currently being used in CWS but also newer solutions (algorithms) being proposed by researchers to better assess the current state of research. We used the following search terms to find papers at the intersection of CWS and algorithms -- "child protective services," "child welfare," "foster care," "child and family services," "algorithm," "computation," "regression," "machine learning," "neural network," "data-driven," "actuarial," "computer program," "application". We used the following inclusion criteria for the articles: 
\begin{itemize}
    \setlength\itemsep{0em}
    \item The paper was peer-reviewed, published work or a systems (or policy) report produced by a government agency.
    \vspace{-0.5mm}
    \item The study (or report) engaged in a technical discussion about the computational methods, predictors and outcomes.
\end{itemize}
Articles that did not meet these two criteria were considered irrelevant for this study and were not included in our review. 
We conducted a comprehensive search to identify relevant research across multiple disciplines. We searched a diverse set of digital libraries which included the ACM Digital Library, IEEE Xplore, Routledge, Elsevier, and Springer. We chose these libraries to take into account  research published in multi-disciplinary conferences and journals. We then cross-referenced the citations of each article to identify  additional articles or government reports that met our inclusion criteria. We did not place any constraints on our search based on the time period in which the papers were published. We identified \textbf{50} relevant articles that met our inclusion criteria.  

\vspace{-0.5em}
\subsection{Data Analysis Approach}
To analyze our data, we conducted a structured qualitative analysis to answer our over-arching research questions. We used a grounded thematic process \cite{braun2006using} to generate codes based on the data as shown in Table \ref{tab:codebook} . We define theory in two ways -- the system discussed in the study was developed using a theoretical framework or the system was developed theoretically based upon factors considered significant in evidence-based social work. The first author coded all of the articles, and co-authors were consulted to form a consensus around codes early in the coding process and again during coding to resolve ambiguous codes. We also coded for descriptive characteristics of the article set as shown in Table \ref{tab:characteristics}. 

\begin{table}[]
    \small
    \centering
    \begin{tabular}{l|c|l}
    \toprule
         \textbf{Code} & \textbf{n} & \textbf{Breakdown} \\
         \hline
         Peer reviewed & 43 & \textbf{40} (social science); \textbf{3} (computer science)\\
         Agency report & 7 & ---\\
         Theory & 5 & \textbf{1} (implemented); \textbf{4} (proposed) \\
         Psychometric scales & 30 & --- \\
         Actual system & 27 & \textbf{15} (RAs); \textbf{11} (PLs); \textbf{1} (MT)\\
         Hypothetical system & 23 & \textbf{13} (RAs); \enspace \textbf{4} (PLs); \textbf{1} (MT); \textbf{5} (S-PL)\\
         Model performance & 35 & ---\\
         \bottomrule
    \end{tabular}
    \vspace{0.1cm}
    \begin{tablenotes}
    \small
        \item[1]\textbf{\hspace{1.1cm} RA}: Risk Assessment model 
        \item[2]\textbf{\hspace{1.1cm} PL}: Placement Recommendation model
        \item[3]\textbf{\hspace{1.1cm} MT}: Child-Foster parent Matching model 
        \item[4]\textbf{\hspace{1.1cm} S-PL}: Characteristics of successful placements
    \end{tablenotes}
    \vspace{0.1cm}
    \caption{Descriptive Characteristics of the Data Set}
    \label{tab:characteristics}
\end{table}

\vspace{-0.3em}
\section{Results}
In this section, we present our key findings from our review of the literature. We begin by first discussing the descriptive characteristics of our data set. Next, we organize and present the results by our three research questions, as shown in Table \ref{tab:codebook}. Finally, we explore the relationship between the computational methods, predictors, and outcomes identified in our analyses.

\begin{table*}[]
    \centering
    \small
    \begin{tabular}{l|l|l|c|c|c}
        \textbf{Research}                    & \textbf{Dimension} & \textbf{Codes} & \textbf{Count} & \textbf{\%} & \textbf{Example}\\
                                      \textbf{Question}& & & & & \\
        \hline
                                       & Inferential Statistics       & Generalized Linear Models (GLM) & 28 & 56\% & \cite{vaithianathan2017developing}\\
\textbf{RQ1}        &                  & Discriminant Analysis/Statistical tests (DAS) & 6 & 12\% & \cite{schwab1989continuum}\\ 
(Computational                         & Machine Learning & Supervised Learning (SUP) & 13 & 26\% & \cite{chouldechova2018case}\\
 Method)       &                       & Unsupervised Learning (UNSUP) & 3 & 6\% & \cite{mcdonald2002predicting}\\
        \hline
        & Demographics                 & Child Demographics (C-DEM) & 20 & 40\% & \cite{andreswari2018preliminary}\\
        &                              & Biological parents Demographics (P-DEM) & 10 & 20\% & \cite{johnson2004effectiveness}\\
        & Systemic Factors             & Characteristics of Agency (AGENCY) & 2 & 4\% & \cite{moore2016assessing}\\
        &                              & Characteristics of Caseworker (WORKER) & 1 & 2\% & \cite{moore2016assessing}\\
        & Child Strengths              & Child Strengths (CHI-S) & 11 & 22\% & \cite{chor2012predicting}\\
        & Child Needs                  & Functioning (CHI-F) & 15 & 30\% & \cite{marshall2000neural}\\
        &                              & Child Behavioral/Emotional Needs (CHI-BE) & 26 & 52\% & \cite{lardner2015restrictiveness}\\
        & Child Risks                  & Suicide Risk (CHI-SR) & 9 & 18\% & \cite{cordell2016patterns}\\
\textbf{RQ2} &                         & Child Risk Behaviors (CHI-BR) & 20 & 40\% & \cite{ringel2018improving}\\
 (Predictor        &                              & Traumatic Experiences (CHI-T) & 30 & 60\% & \cite{baird1999risk}\\
variables)      
        &                              & Child Involvement in CWS (CHI-CWS) & 9 & 18\% & \cite{vaithianathan2017developing}\\

        & Bio-Parent Risk/Needs        & Needs and Risky behavior (PAR-NS) & 26 & 52\% & \cite{chor2012predicting}\\
        
        & Foster Parents               & Characteristics (income, occupation) (FP-CHAR) & 4 & 8\% & \cite{andreswari2018preliminary}\\
        &                              & Preferences (FP-PREF) & 2 & 4\% & \cite{moore2016assessing}\\
        &                              & Past performance (FP-PAST) & 1 & 2\% & \cite{moore2016assessing}\\
        &                              & Capabilities (training/certifications) (FP-CAPS) & 1 & 2\% & \cite{moore2016assessing}\\
        \hline
        & Outcome                      & Risk of a future maltreatment event (RISK) & 28 & 56\% & \cite{johnson2004effectiveness}\\
\textbf{RQ3} &                         & Placement recommendation for a child (PLACE) & 15 & 30\% & \cite{chor2012predicting} \\
 (Outcome       &                              & Matching children with foster parents (MATCH) & 2 & 4\% & \cite{moore2016assessing}\\
 Variables)        &                              & Characteristics of a successful placement (S-PLACE) & 5 & 10\% & \cite{stone1983prediction}\\
                     
    \end{tabular}
    \vspace{0.2cm}
    \caption{Structured Codebook: Dimensions are mapped onto their respective research questions}
    \label{tab:codebook}
\end{table*}

\vspace{-0.5em}
\subsection{Descriptive Characteristics of the Data Set}
The majority of the papers (n=40 or {80\%}) were published in social science venues with 3 papers (6\%) published in computer science conferences, \cite{aburas2018child, andreswari2018preliminary, chouldechova2018case} all in 2018. We also included 7 reports (14\%) from non-profit organizations, including the Children's Research Center \cite{CRC}. One study  discussed an algorithm which was theoretically constructed based on child-welfare research literature and four studies proposed theoretically-driven solutions. 30 papers (60\%) employed psychometric scales \cite{furr2017psychometrics} to assess the strengths, needs and risks associated with foster children and/or the biological parents. 27 papers (54\%) discussed an actual algorithmic system and 23 papers (46\%) proposed a new algorithmic system. Model performance was reported by 35 papers (70\%).

\subsection{Computational Methods used to build Algorithms (RQ1)}
In this section, we discuss the computational methods used to develop algorithms and organize them into the \textit{Inferential Statistics} and \textit{Machine Learning} dimensions.

\subsubsection{Inferential Statistics approaches}
\begin{figure}
  \includegraphics[scale=0.50]{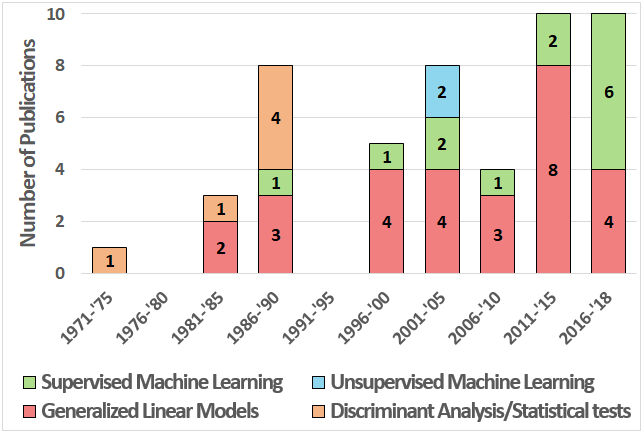}
  \caption{Methods used to build Algorithms (RQ1)}
  \label{fig:compmodels}
\end{figure}

Inferential statistics account for computational methods used in the majority of papers (68\%), with 28 papers (56\%) using a form of a generalized linear model (GLM). In Figure \ref{fig:compmodels}, we see a dramatic rise in the use GLMs after 1995, i.e., the post-\emph{SACWIS} era. GLMs are being used to develop mostly two types of models; actuarial risk assessment and placement recommendation models. There was a general trend around the use of GLMs for developing risk assessment models \cite{vaithianathan2017developing, camasso2013decision, fluke2010placement}. We also identified two major concern surrounding GLMs: their atheoretical and reductive nature and performance with respect to outliers. 

Social scientists use validated psychometric scales \cite{lyons2004measurement, camasso1995prediction} to quantify the level of risk. GLMs have been developed using these psychometric scales, such as the CANS Algorithm \cite{chor2012predicting} that only uses the most statistically significant items from the scale. This reductive and atheoretical model development has received criticism \cite{camasso2013decision, schwalbe2008strengthening, shlonsky2005next}. Each case of child neglect/abuse is contextually different and factors that are significant for one case might be peripheral to another. Moreover, GLMs do not account for the contextual factors that influence caseworker decisions leading to variable omission bias \cite{camasso2013decision}.

%Risk assessment tools are commonplace in social sciences. There are several validated psychometric scales that social scientists use to quantify the level of risk such as CANS (Child and Adolescents Needs and Strength) \cite{lyons2004measurement} and WRAM (Washington Risk Assessment Matrix) \cite{gambrill2000risk} assessment scales. GLMs have been developed using these psychometric scales, such as the CANS Algorithm \cite{chor2013overview} that only uses the most statistically significant items from the scale. This reductive and atheoretical model development approach has received criticism \cite{camasso2013decision, schwalbe2008strengthening, shlonsky2005next}. Each case of child neglect/abuse is contextually different and characteristics that are significant for one case might be peripheral to another. Furthermore, GLMs do not account for the contextual factors that influence caseworker decisions leading to variable omission bias \cite{camasso2013decision}. Studies have recommended using cumulative risk modeling that accounts for accumulation of risk \cite{mackenzie2011toward} and using the ROC (Receiver Operating Characteristic) curve to improve predictive validity \cite{camasso2013decision}. However, they still do not address the core of the problem founded in the reductive and atheoretical nature.

Outliers can significantly impact the performance of a regression model \cite{ott2015introduction}. Traditionally, regression models seek to omit outliers as a means of improving predictive power and still account for majority of the variance explained by significant variables \cite{ott2015introduction}. However, for CWS, cases of severe abuse and neglect are the statistical outliers \cite{brand1999behavior}. Regression models that are designed to predict the most moderate (average) outcomes tend to perform poorly on outliers \cite{stevens1984outliers}. In terms of CWS, poor performance on outliers raises several ethical and accountability concerns \cite{corrigan_2019}. 

Four papers (8\%) used discriminant analysis to differentiate between the characteristics of foster children served by different placement settings. Figure \ref{fig:compmodels} illustrates that discriminant analysis was a popular technique during 1985-1990, however, with the advent of regression techniques it gradually faded away. These papers were some of the earliest attempts at introducing algorithms to aid decision-making in CWS. However, the data was limited and its quality questionable because of the lack of standardized data collection processes \cite{sicoly1989prediction}.

\vspace{-1mm}
\subsubsection{Machine Learning (ML) approaches}
Machine Learning methods in CWS gained some momentum as early as 1986 with the introduction of \textit{PLACECON}, a system designed to assist CWS with placement decisions \cite{schuerman1986computer}. However, with the increasing popularity of risk assessment models and limited funding available, resources were directed towards traditional regression models. Figure \ref{fig:compmodels} shows a resurrection of ML methods starting 2015 and a growing interest within the computer science communities towards studying the research problems in CWS starting 2018 \cite{aburas2018child, andreswari2018preliminary, chouldechova2018case, harrison2018tale}. Thirteen papers (26\%) utilized ML methods in the form of decision trees, Bayesian networks or inference trees. Decision-tree learning has been popular as a means of organizing large amount of factual and empirical knowledge in the form of rules \cite{sicoly1989computer}. The CART (Classification and Regression Trees) algorithm has been recently used to build a child-foster parent matching system \cite{andreswari2018preliminary}. It has also been used to identify the characteristics of the most troubled children in CWS \cite{cordell2016patterns} as well as to study trends in child abuse and neglect data \cite{aburas2018child}. However, with such a strong emphasis on risk assessment, Children's Research Center \cite{CRC} used ML methods to develop the Structured Decision-Making (SDM) model. 

SDM is a decision-making framework where a risk assessment tool is used in conjunction with clinical assessment \cite{johnson2004effectiveness}. SDM utilizes an array of ML tools such as decision, value and inference trees, and Bayesian networks \cite{gregory2012structured} and has been adopted by CWS in several states \cite{bosk2018counts}. However, several studies have also shown that SDM produces mixed results especially when accounting for race and ethnicity \cite{d2008risk, danktert2013risk, johnson2003california}. There is also an ongoing struggle between caseworker's theoretical assessments and the tool's empirical judgment \cite{schwalbe2008strengthening, shlonsky2005next}. Three papers (6\%) used unsupervised ML methods in the form of neural networks \cite{marshall2000neural, mcdonald2002predicting} and natural language processing (NLP) \cite{brindley2018can}. Brindley et al. \cite{brindley2018can} propose a web platform that allows foster youth to create personalized goals and talk to a chat-bot that uses NLP to parse inputs and respond intelligently with recommendations about goals, finances, and housing. McDonald et al. \cite{mcdonald2002predicting} and Marshall et al. \cite{marshall2000neural} propose the use of neural networks over regression techniques because their non-parametric approach performs better at modeling non-linear relationships and interactions. 

One possible reason for the perpetual conflict between ML risk assessment tools and caseworkers' assessment might be at the core of Machine Learning itself and how it handles outliers. Statistical outliers in the case of child maltreatment are the most severe cases of child abuse and neglect \cite{brand1999behavior}. Researchers \cite{bakar2006comparative} suggest that in the case of CWS, outliers are often more important for caseworkers and demands significant attention beyond the norm. Figure \ref{fig:compmodels}, depicts a significant dearth in the use of unsupervised learning methods with only two papers published in early 2000s \cite{mcdonald2002predicting, marshall2000neural} and one paper published in 2018 \cite{brindley2018can}. Employing neural networks in social sciences comes with its own complexities because there needs to be transparency about the proposed decisions \cite{chouldechova2018case}. Vaithianathan et al. \cite{vaithianathan2017developing} explored several ML methods such as Naive Bayes and Random Forests for risk assessment and achieved higher accuracies. However, they reverted to using a probit regression model because the outcomes were more explainable.

\subsection{Predictors used in Algorithms (RQ2)}
In this section, we examine the predictors that are being used in algorithms in CWS. Most algorithms are using over a hundred predictors so we systematically coded them and then grouped the emergent codes into seven dimensions (see Table \ref{tab:codebook}).

%\vspace{-0.5em}
\subsubsection{Demographics and Systemic Factors}
Child demographics were accounted for by 20 papers (40\%) and biological parents demographics were accounted for by 10 papers (20\%). Surprisingly, more than half the papers did not include child or parent demographics in their models even though racial and ethnic disparities in CWS have been recognized in social sciences \cite{needell2003black, putnam2013racial, dettlaff2011disentangling}. Figure \ref{fig:predictors}, illustrates that after 1990, there was a decline in the number of studies that used demographic variables in their algorithms. The \textit{Systemic factors} dimension includes factors associated with CWS, such as, characteristics of the agency and caseworkers. Two papers (4\%) use variables relating to characteristics of the agency, such as location and staffing vacancies and one paper (2\%) accounted for the characteristics of the caseworker, such as, caseloads and the level of training. This is surprising because child-welfare literature acknowledges the impact caseworkers have on child outcomes \cite{ryan2006investigating, carnochan2013achieving}. The caseworker is a child's primary contact between the biological parents, foster parents and CWS. They navigate through the system and find services for children and families. In fact, caseworker turnover is directly associated with placement instability \cite{carnochan2013achieving}. Factors that lead to high caseworker turnover include low salary, high caseloads, administrative burdens, low levels of training and lack of supervisory support \cite{carnochan2013achieving}. Systemic factors is one of the biggest reasons why children experience multiples placement moves in CWS \cite{cross2013children}. This once again alludes to the atheoretical model construction that does not account for the salient factors well-established in evidence-based social work.  

%\vspace{-0.5em}
\subsubsection{Foster-child related factors}
\begin{figure}
  \includegraphics[scale=0.5]{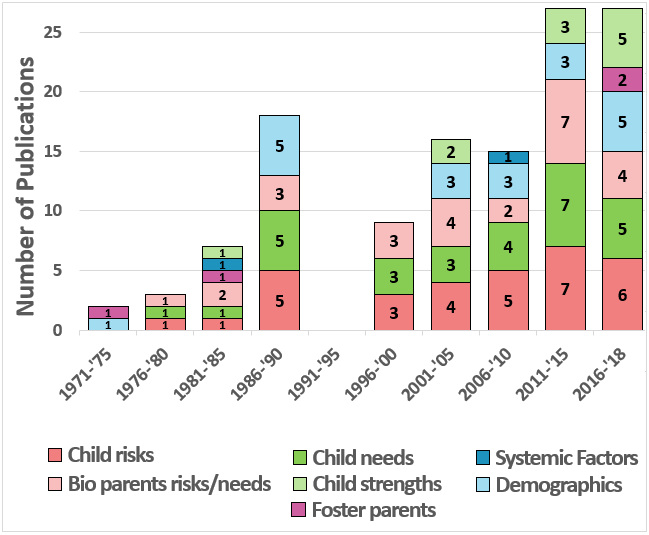}
  \caption{Predictors used in Algorithms (RQ2)}
  \label{fig:predictors}
  \vspace{-1mm}
\end{figure}

Seven codes emerged out of the coding process and were grouped into three dimensions: child strengths, child needs, and child risks. 11 papers (22\%) use variables that align with \textit{Child Strengths}, such as, interpersonal skills, coping skills and level of optimism. Twenty-six papers (52\%) took into account a child's emotional and behavioral needs and 15 papers (30\%) recorded the child's day-to-day well-being and functioning, such as, their school attendance and behavior, personal hygiene and communication skills. We also coded for variables associated with risk factors that endanger child well-being. Suicide risk, risk behaviors, traumatic experiences, and child involvement with CWS were our four emergent codes that were grouped under the \textit{Child Risks} dimension. 9 papers (18\%) conducted a mental health screening to see if a child was suicidal or having suicidal thoughts. 20 papers (40\%) accounted for risk behaviors such as self-harm, recklessness, social misbehavior, and 30 papers (60\%) accounted for traumatic experiences such as neglect, physical/sexual abuse, history of family violence, and community violence. We noticed a trend here in that, almost all the risk assessment systems focused heavily on the \textit{Child Risks} dimension, whereas, placement recommendation systems focused on the \textit{Child Needs} dimension. Figure \ref{fig:predictors} depicts a rise in the number of studies that account for child strengths, child risk and child needs since 1995, that is, the post-\emph{SACWIS} era. This alludes to the fact that these child characteristics are well-documented in \emph{SACWIS} and are being used for modeling purposes. 

All the studies we reviewed accounted for foster child related factors in terms of their needs and associated risks. However, only one study accounts for the child's interactions with other people, such as siblings, relatives, and the system itself. Moore et al. \cite{moore2016assessing} account for factors such as \textit{Placement with a sibling}, \textit{Proximity to child's home/relatives}, and \textit{Characteristics of the agency and caseworker}; factors well-studied in child-welfare literature \cite{carnochan2013achieving}. Fluke et al. \cite{fluke2010placement} found that placement decisions may be made as a result of interaction effects of non-case related factors such as characteristics of the agency and/or the caseworker. A study conducted in San Diego County found that 70\% of the placement moves were a result of systemic or policy related factors \cite{james2004foster}. 

\subsubsection{Biological parents related factors}
26 papers (52\%) accounted for the biological parents' risk behaviors and needs, such as, physical/mental health, substance abuse problems, residential stability and knowledge of child's needs. We coded these variables into the \textit{Bio-Parents Risks/Needs} dimension. Figure \ref{fig:predictors} shows that biological parent factors have been consistently used by several studies, however, we see a decline during 2005-2010. We also see a rise in the use of foster child related factors during the same time period. The introduction of CANS algorithm that focuses on the child's level of need may be a plausible explanation for this trend. Different algorithms are using biological parent related variables differently. For example, risk assessment models quantify biological parents' risky behavior so as to discern the risk of a future maltreatment. On the other hand, placement recommendation models are using this dimension to determine the level of trauma a child has experienced and recommend a placement setting based on their level of need. Factors surrounding biological parents have been studied in great detail and accounted for by most algorithms. 

\vspace{-0.3em}
\subsubsection{Foster parents related factors}
Four papers (8\%) that we reviewed accounted for the characteristics of the foster parents, that is, their income level, occupation, demographics etc. Figure \ref{fig:predictors} shows that only 4 studies account for foster parent related factors with a significant gap between 1985 and 2016 where no study accounted for these factors. Two papers (8\%) look at the preferences of foster parents and one paper (2\%) accounts for the foster parents' past performance and capabilities. Matching children with foster parents that are trained and prepared to meet their behavioral and emotional needs leads to increased stability for the children \cite{carnochan2013achieving}. Matching children with foster parents that come from the same cultural background also leads to better outcomes because it leads to smoother transitions, lower stress and a feeling of security for the children \cite{brown2009benefits}. These factors are well-studied in child-welfare literature \cite{carnochan2013achieving, redding2000predictors, webster2000placement}, however, we see that very little research has been done from an algorithmic perspective. CWS has historically focused on ensuring safety and permanency rather than child well-being, that is, improving the quality of lives of foster children \cite{berger2009estimating}.

\vspace{-0.3em}
\subsection{Target Outcomes of Algorithms (RQ3)}
In this section, we examine the target outcomes of the algorithms used in CWS. Figure \ref{fig:outcomes} depicts the trends in the target outcomes that algorithms have sought to model. 

\vspace{-0.3em}
\subsubsection{Risk Assessment}
\begin{figure}
  \includegraphics[scale=0.51]{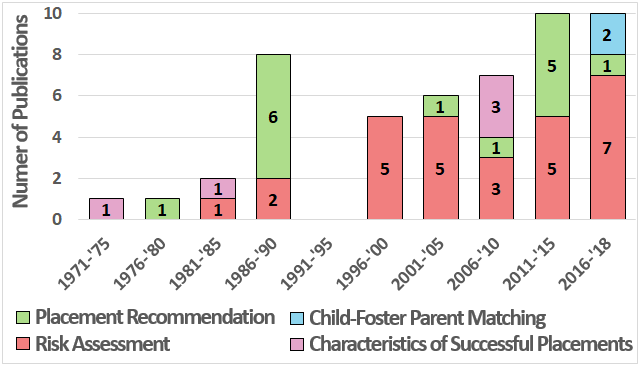}
  \caption{Target Outcomes of Algorithms (RQ3)}
  \label{fig:outcomes}
\end{figure}

Predicting the risk of future maltreatment involves developing models using the empirical study of cases of child abuse/neglect \cite{baird1999risk}. The factors that show a strong association with abuse and/or neglect outcomes are selected to create an actuarial model which is then used to assess new cases of alleged abuse/neglect. Twenty-eight papers (56\%) focused on predicting risk as their target outcome. Figure \ref{fig:outcomes} illustrates that risk assessment has always received significantly more attention than any other outcome since the introduction of regression models in social sciences. The greatest criticism against these models is that they are not theoretically founded; these models are probabilistic in nature and not causal \cite{baird1999risk, koh2008propensity, schwalbe2008strengthening, shlonsky2005next}. Therefore, these models need to be empirically validated by follow-up studies to ensure their reliability. Direct comparison of any two actuarial models is a hard problem \cite{baird1999risk} and requires an in-depth understanding of the contexts in which the predictors were collected, measured and weighted in the models.
%This supposedly allows caseworkers to use a structured, standardized and empirical methodology to inform the decision-making process. 

Studies conducted on risk assessment models show that these models are more accurate at predicting target events like child maltreatment than unaided judgment, however, they lack utility \cite{schwalbe2008strengthening}. Seven papers (14\%) discuss the Structured Decision-Making (SDM) model, a framework that integrates predictive and contextual assessments. CWS in several states have developed their own versions of SDM, however, there are significant enough differences to treat them independently as part of our review. Even though SDM is designed to assist caseworker decisions, studies have found that there are constant disagreements between the tool (empirically-driven) and caseworker assessment (conceptually/theoretically-driven) to the point that caseworkers detest using these tools as they were intended \cite{schwalbe2008strengthening}. However, caseworkers must continue to rely on these tools as a means of standardizing decisions in CWS, especially in cases of high uncertainty \cite{shlonsky2005next}.

\vspace{-0.2em}
\subsubsection{Placement Recommendations and Successful Placements}
Models that focused on these two target outcomes were the precursors in the development of algorithms in CWS. Figure \ref{fig:outcomes} depicts that these target outcomes were being studied during the time period of 1985-1990. However, no studies were published between 1991 and 2005 that focused on these target outcomes. A plausible explanation for this decline would be the increased focus on studying risk assessment during that period. 20 papers (40\%) discussed recommendation systems for foster care placements. 

\vspace{-0.1em}
The most prominent algorithm that determines the placement criteria based on a child's level of need is the CANS algorithm \cite{chor2012predicting}. 6 papers (12\%) discuss the CANS algorithm which is developed using the CANS psychometric scale \cite{lyons2004measurement}. CWS in a few states have developed their own versions of this algorithm, and therefore, were treated independently as part of our review. It makes a recommendation from six levels of care in the order of increasing severity -- independent living, transitional living program, foster home, specialized foster care, group home, and residential treatment center. It is used in a hybrid approach in conjunction with a multi-disciplinary team which allows CWS to follow a standardized admission criteria for cases with lower levels of uncertainty \cite{chor2012predicting}. This is a good initial approach to ensure child safety, however, it is a minimal approach and does not seek to improve the quality of a child's life. 

%There is high concurrence between the approaches at the two ends of the placement spectrum, that is, the least and the most restrictive settings. This implies that the algorithm and the multi-disciplinary team tend to agree on the least and most severe cases of child maltreatment. But there are discordant recommendations towards the center where there is higher uncertainty in measuring a child's level of need. However, these disagreements, that is, recommendation made by the team versus the recommendation made by the algorithm do not differ by more than one level of care (on a scale of six) \cite{chor2013patterns}.

\begin{table}[]
    \centering
    \small
    \begin{tabular}{>{\raggedright}p{1cm}|>{\raggedright}p{0.6cm}|>{\raggedleft}p{1.1cm}>{\raggedleft}p{1.2cm}>{\raggedleft}p{1cm}>{\raggedleft\arraybackslash}p{1.5cm}}
    \toprule
             & & \multicolumn{4}{c}{\textbf{Computational methods}} \\\cline{3-6}
             \cr %& & & & & &
             & &  Supervised Machine Learning & Unsupervised Machine Learning & Generalized Linear models & Discriminant analysis/ statistical tests\\
        \hline
        & RA    &  9 & 2 & 16 & 1 \\
        \textbf{Outcome} & PL   &  2 & 1 & 8 & 4 \\
        \textbf{Variables} & MT   &  2 & - & - & -\\
        & S-PL &  - & - & 4 & 1\\
        \bottomrule
    \end{tabular}
    \vspace{0.1cm}
    \begin{tablenotes}
    \small
        \item [1]\textbf{RA}: Risk Assessment model 
        \item[2]\textbf{PL}: Placement Recommendation model
        \item[3]\textbf{MT}: Child-Foster parent Matching model 
        \item[4]\textbf{S-PL}: Characteristics of successful placements
    \end{tablenotes}
    \vspace{0.1cm}
    \caption{Crosstabs between Methods and Outcomes}
    \label{tab:R1vR3}
\end{table}

\vspace{-0.1em}
\subsubsection{Child-Foster Parent Matching}
This approach seeks to match the specific needs of a child with the capabilities of foster parents. That is, placing children with foster parents who are trained and certified to manage their needs. It is a proactive approach towards improving the quality of lives of children and not just minimizing risk of maltreatment. Figure \ref{fig:outcomes} shows that \textit{Child-Foster parent matching} has only been implemented since 2015 (2 studies). This approach is different from the placement recommendation approach in that it addresses the specific needs of the child and preferences of the caregiver. For instance, matching with respect to child temperament, parent temperament, and parental expectations leads to increased stability \cite{redding2000predictors}. Placing children who have higher emotional needs with foster parents who prefer to be emotionally involved offers these children a better chance towards stability \cite{walsh1990studies} than placing these children in a restrictive treatment setting. \textit{Child-Foster parent matching} is well-studied in evidence-based social work and is known to improve stability and permanency outcomes \cite{carnochan2013achieving, redding2000predictors}. However, there is a dearth of information within CWS on how to guide this process \cite{redding2000predictors}. This is a significant knowledge gap for both CWS and social scientists who seek to computationally model this approach. Moore et al. \cite{moore2016assessing} recently validated a matching algorithm that was implemented by CWS in the state of Kansas for resulting in more stable placements. 

\vspace{-0.3em}
\subsection{Relationship between Methods, Predictors and Outcomes}

\begin{table*}[]
    \centering
    \small
    %\begin{tabular}{l|l|cccc|cccc}
  \begin{tabular}{>{\raggedright}p{1.1cm}|>{\raggedright}p{3.5cm}|>{\raggedleft}p{0.8cm}>{\raggedleft}p{0.7cm}>{\raggedleft}p{0.7cm}>{\raggedleft}p{1cm}|>{\raggedleft}p{0.8cm}>{\raggedleft}p{0.8cm}>{\raggedleft}p{0.8cm}>{\raggedleft\arraybackslash}p{1.2cm}}
                 \toprule
                  & & \multicolumn{4}{c}{\textbf{Outcome Variables}} & \multicolumn{4}{c}{\textbf{Computational Methods}} \\\cline{3-10} 
                  \cr %& & & & & & & & & &
                  & & RA & PL & MT & S-PL & SUP & UNSUP & GLM & DAS\\
                  \hline
                  & Child demographics & 8 & 6 & 2 & 4 & 7 & 1 & 7 & 5 \\
                  & Bio-parents demographics & 5 & 2 & 2 & 1 & 5 & 1 & 2 & 2\\
                  & Characteristics of Agency & - & - & - & 2 & - & - & 2 & -\\
                  & Characteristics of Caseworker & - & - & - & 1 & - & - & 1 & -\\\cline{2-10}
                  
                  & Child Strengths & 3 & 5 & 2 & 1 & 5 & 1 & 5 & -\\
                  & Functioning & 3 & 10 & 2 & - & 5 & 1 & 7 & 2\\
    \textbf{Predictors}    & Child Behavioral/Emotional Needs & 7 & 13 & 2 & 4 & 7 & - & 13 & 6\\
                  & Suicide Risk & 2 & 7 & - & - & 3 & - & 4 & 2\\
                  & Child Risk Behaviors & 7 & 12 & 1 & - & 6 & 1 & 10 & 3\\
                  & Traumatic Experiences & 15 & 10 & 2 & 3 & 9 & 1 & 16 & 4\\
                  & Child Involvement with CWS & 2 & 6 & 1 & - & 3 & - & 3 & 3\\\cline{2-10}
                  
                  & Bio-Parent Risk/Needs & 15 & 8 & - & 3 & 7 & 1 & 16 & 3\\\cline{2-10}
                  
                  & Foster parent characteristics & - & - & 2 & 2 & 2 & - & 1 & 1\\
                  & Foster parent preferences & - & - & 1 & 1 & 1 & - & - & 1\\
                  & Foster parent past performance & - & - & 1 & - & 1 & - & - & -\\
                  & Foster parent capabilities & - & - & 1 & - & 1 & - & - & -\\
                  \bottomrule
    \end{tabular}
    \vspace{0.1cm}
    \begin{tablenotes}
    \small
        \item [1]\textbf{\hspace{6.3cm} RA}: Risk Assessment model \hspace{21mm} \textbf{SUP}: Supervised Machine Learning 
        \item[2]\textbf{\hspace{6.3cm} PL}: Placement Recommendation model \hspace{8mm} \textbf{UNSUP}: Unsupervised Machine Learning
        \item[3]\textbf{\hspace{6.3cm} MT}: Child-Foster parent Matching model \hspace{0.6cm} \textbf{GLM}: Generalized Linear models
        \item[4]\textbf{\hspace{6.3cm} S-PL}: Characteristics of successful placements \hspace{0mm} \textbf{DAS}: Discriminant Analysis/Statistical tests
    \end{tablenotes}
    \vspace{0.1cm}
    \caption{Relationship between Computational Methods (RQ1), Predictors (RQ2) and Outcomes (RQ3)}
    \label{tab:Big_crosstabs}
\end{table*}

\subsubsection{Relationship between Algorithms (RQ1) and Outcomes (RQ3)}
Table \ref{tab:R1vR3} depicts crosstabs between the computational methods used and the outcome from all the papers in our corpus. We saw that generalized linear models have mostly been used for developing risk assessment models (16 studies) followed by placement recommendation models (8 studies). Even with the emergence of newer machine learning methods, majority of the studies still continue to focus on risk assessment. 9 studies used supervised machine learning for risk assessment, 2 studies focused of placement recommendation and 2 studies focused on child-foster parent matching.

\vspace{-0.3em}
\subsubsection{Relationship between Predictors (RQ2) and Outcomes (RQ3)}
Table \ref{tab:Big_crosstabs} depicts the crosstabs between predictors used by computational models and the outcome they seek to predict. First, we cross examine the risk assessment models with respect to the predictors that inform child characteristics. Majority of the models use a combination of predictors that assess \textit{Child Behavioral/Emotional Needs} (7 studies), \textit{Child Risk Behaviors} (7 studies), and \textit{Traumatic Experiences} (15 studies). Several predictors coded under these dimensions (e.g., self-harm, recklessness, physical/sexual abuse) are assessed by a caseworker at an initial investigation and made available for predictive modeling. These predictors might already exist in the data if the family has previously come under the attention of CWS. This approach of aggregating the negative aspects of people's lives while ignoring the positive aspects has been criticized because of its deficit-based nature \cite{brown2019toward}. There is also an overlap between the \textit{Traumatic Experiences} of a child and the \textit{Bio-Parents Needs/Risk Behavior} because the same predictors (for e.g., history of physical/sexual abuse, medical trauma, parents' criminal activity) are used to conduct both \textit{needs} assessment for a child and \textit{risks} assessment for a parent.

Next, we cross examine the predictors used by placement recommendation models. These models are not employed at the onset of an investigation and are used by CWS when a child needs to be placed in a permanent foster care setting. These models are generally more equitable as compared to risk assessment models because they try to weigh-in the positive characteristics of a child, such as talents, interests, cultural identity, and school achievements to find an appropriate placement setting that meets their needs. Table \ref{tab:Big_crosstabs} shows that placement recommendation models account for predictors around \textit{Child Strengths} (5 studies), \textit{Functioning} (10 studies), and \textit{Child Behavioral/Emotional Needs} (13 studies) to weigh in the positive aspects and needs of a child and balance that with predictors around \textit{Child Risk Behaviors} (12 studies) and \textit{Traumatic Experiences} (10 studies) to find a suitable placement setting well equipped to meet their needs.

\vspace{-0.3em}
\subsubsection{Relationship between Methods (RQ1) and Predictors (RQ3)}
Table \ref{tab:Big_crosstabs} depicts the crosstabs between predictors and computational methods. Most computational methods including both supervised machine learning and generalized linear models focused on \textit{Child Behavioral/Emotional Needs}, \textit{Child Risk Behaviors}, and \textit{Traumatic Experiences} to assess the risk of a maltreatment event or the needs of a child. Some predictors that inform these three codes include traumatic events (e.g., physical/sexual abuse, medical trauma), child's conduct and anger management, and delinquent behavior. After an initial investigation is conducted by a caseworker and psychometric risk assessments completed, these predictors become available for modeling. However, quantifying risk from such a narrow set of predictors has been criticized because it fails to account for the wide range of risk factors that arise as a result of systemic issues in CWS itself \cite{gambrill2000risk}.

\vspace{-0.3em}
\section{Discussion}
%In this section, we discuss the implications of our findings and future research directions.

\subsection{Algorithms Need to be theoretical \& context-aware (RQ1)}
Overall, we found a lack of theoretically derived and validated algorithms that demonstrated that they took measures to integrate knowledge from the social sciences into their designs. Only one study \cite{moore2016assessing} constructed their model based on the child-welfare literature. Four studies even discussed this lack of theory and proposed solutions in the form of cumulative risk models \cite{mackenzie2011toward}, causal models \cite{schwalbe2004re}, and revised SDM models grounded in risk and resilience theory \cite{schwalbe2008strengthening}. Yet, based on the published literature, such models have yet to be consistently implemented.   

This finding is problematic because it shows that these algorithms ignore many  factors that affect how decisions are made in CWS. For instance, the decisions are often constrained by current policies or the scarce resources \cite{gambrill2001need}. Many current empirical models frustrate child-welfare workers because they do not account for such systemic factors. While some researchers have suggested \cite{kleinberg2015prediction} that empirical prediction is enough and that theory, context, or causal inferences are not always necessary in policy making when outcomes remain desirable, we argue that this is not a desirable stance to take in child-welfare contexts because there is significant debate on how and which types of data, models, and outcomes are to be used in predictive modeling (with or without theory). Empirical knowledge related to child-welfare practice is fragmented and social science theories must be used to fill the gaps \cite{gambrill2000risk}.

Therefore, we recommend that human-centered theoretical approaches be used to incorporate factors arising from evidence-based social work \cite{carnochan2013achieving} and understand the causal pathways that often dictate decision-making processes. Such algorithms that are informed by appropriate causal theory would also have a greater likelihood of utilization as compared to their a-theoretical counterparts \cite{schwalbe2004re}. Significant work has also found disconnects between the functioning of algorithms and their social interpretations \cite{baumer2017toward}. We see a similar phenomena in CWS where the caseworkers using Structured Decision Making (SDM) model must translate information from both forms of assessments (clinical and algorithmic) leading to uncertainty and unreliable decision making \cite{shlonsky2005next}. Therefore, algorithms that are meant to aid decision-making often become the source of frustration and force caseworkers to abandon their contextual judgments \cite{schwalbe2008strengthening}. Human-centered theoretical approaches can help by placing the meaning-making process \cite{pink2013applying} at the center of the design process. It can help designers understand the theory of practice and uncover practitioners' sense-making processes (e.g., how they perceive quantified metrics \cite{baumer2017toward}). Child-welfare workers are generally not trained in statistical thinking and make decisions based on experience, intuition, and individual heuristics \cite{gambrill2000risk}. Human-centered theoretical approaches can help us understand the mental models of child-welfare workers, inform feature selection (design), as well as interpret the results (evaluation).  

Our results also indicate that several states adopted the SDM approach because it was supposed to integrate predictive and contextual assessments, however, it has fallen short of that goal \cite{shlonsky2005next, schwalbe2008strengthening}. There are several factors at play in regards to any child-welfare case and it becomes critical to offer context to the case instead of focusing on a few broad factors without giving weight to important nuances \cite{Report-LA-SDM}. For instance, understanding contextual knowledge with respect to an organization requires incorporating the organizational memory of the organization and its people \cite{march1991exploration, ackerman2004organizational} which is inherently HCI research. Social workers are trained in writing detailed case notes by translating their context-specific experiences into text \cite{clifford1986writing}. This unstructured, unanalyzed, textual data is added to \textit{SACWIS} systems \cite{us2017afcars}. We hypothesize that valuable theoretical signals from these case notes can be considered within methodological approaches like topic modeling that can make good use of such unstructured data. Indeed, in recent years, HCI has developed a rich methodological tradition \cite{baumer2017comparing, muller2016machine, chancellor2016quantifying} of using signals from such unstructured data as predictors within algorithms to study complex, sociotechnical systems.

%HCI research on organizational memory \cite{march1991exploration,ackerman2004organizational} can be used in conjunction with algorithmic approaches. Drawing on the work of Mark Ackerman on organizational memory, the complex contextual understanding of the members of an organization can be stored and reused as the knowledge base of that organization \cite{ackerman2004organizational}.

\vspace{-2mm}
\subsection{Going Beyond what is "Easily Quantifiable" (RQ2)}
Our results suggest that majority of the algorithms used predictors around child and parent characteristics, such as their needs, strengths, and associated risks  (see Table \ref{tab:codebook}). The vast majority of these predictors that are used for predictive modeling are derived from information that is easily available and readily quantifiable. For example, child-welfare workers use psychometric scales \cite{d2008risk} to assess child and parent associated risks and needs during an initial investigation which then become available for predictive modeling. Some of these predictors are found in almost every risk assessment model even though they have no predictive validity. For instance, severity of abuse is easily quantifiable and is found in several risk assessment models even though there is little to no indication that it is related to recurrence of abuse \cite{camasso2000modeling}. Moreover, several predictors (parenting skills, parent conflict etc.) have not been properly validated \cite{gambrill2000risk}, and can lead to unreliable predictions \cite{gambrill2000risk}. Such issues led to the Illinois CWS (in 2017) to shut down their predictive analytics program \cite{dcf_illinois}. In addition, none of the predictors account for the temporality of risk assessment. After an allegation of abuse, the assumption of escalation is the baseline for risk assessment leading to inflated risk scores and excessive interventions from CWS \cite{simpson2000statistical}. 

Human-centered theoretical approaches can result in a rigorous feature selection process that relies on predictors that have been well-studied, understood, and validated in social sciences \cite{baumer2017toward}. De Choudhury et al.’s \cite{de2018integrating} work in mental health is a good example, where the researchers validated constructs, focused on data biases and unobserved factors, as well as conducted sensitivity analysis. Moreover, it compels us to look towards sources of information that have been hereto hard to quantify. For instance, referring to our prior example around case notes, advances in natural language processing \cite{wallach2006topic} now allows us to quantify and make holistic inferences about all the stakeholders involved in a child-welfare case. This can address persistent issues among cases which appear similar based on the empirical data but exhibit high variation in outcomes \cite{gambrill2000risk}. 

Furthermore, human-centered participatory design \cite{baumer2017toward} allows for HCI researchers to actively engage with domain experts in child-welfare to understand how risk accumulates (and how to model it), as well as engage with other stakeholders to better understand the systemic factors around policies, laws and organizational culture \cite{whittaker2017child}. Here, PD \cite{muller2009participatory} can navigate the thorny, contextual differences between different legal and policy systems and needs/values of stakeholders. Lodato and DiSalvo \cite{lodato2018institutional} highlight the different forms and limitations of PD as well as how PD can be conducted within such institutional constraints. Advances in CWS data systems \cite{acf_ccwis} can accommodate for the collection of several new predictors concerning child well-being and systemic factors. Here, the active consideration of needs and values of all stakeholders can help avoid the same reliability and validity pitfalls for the new predictors that exist for many of the current predictors.

\vspace{-0.5em}
\subsection{Improve Lives and not just `Minimize Risk'(RQ3)}
One of the fundamental goals of CWS in the United States is to ensure positive outcomes for foster children \cite{HowCWSworks}, however, as our results confirm, majority of the efforts in computational modeling continues to be focused on risk assessment (see Table \ref{tab:codebook}). Risk assessment models only seek to minimize the risk of future harm and not improve the quality of lives of foster children. The target outcome of "risk of maltreatment" is poorly defined \cite{zuravin1999child}. Federal and State law dictate how child abuse and neglect are defined and the state definitions often vary and establish the grounds for intervention by CWS \cite{HowCWSworks}. Algorithms are trained on cases of substantiation, that is, cases where CWS judged maltreatment to have occurred \cite{gambrill2000risk}. This judgment in itself is very subjective and depends on state laws, policies, and CWS intervention criteria which is often dictated by the level of funding and caseloads \cite{carnochan2013achieving}.

Human-centered approaches can help theoretically define not only the predictors but also the target outcomes with the help of stakeholders and domain experts to ensure these key ingredients needed for algorithm design are validated and reliable. Human-centered participatory design can also unravel concerns around the social interpretations of algorithmically-based systems. For instance, Brown et al \cite{brown2019toward} investigated the community perspectives of risk assessment models in child-welfare. Child-welfare workers criticized these models because of their `deficit-based' nature, that is, this approach only captures negative inputs to predict a negative outcome. There is growing concern that such an approach drives disproportionate negative caseworker perceptions that ultimately leads to negative actions \cite{brown2019toward}. Badillo-Urquiola et al. \cite{badillo2018chibest} and Pinter et al. \cite{pinter2017adolescent} also recognized the problems with a deficit-based framing in that it creates a sense of moral panic and diverts attention away from positive outcomes. They suggested that researchers focus on "strength-based approaches" that focus on positive factors that help improve lives. 

CWS should actively focus on approaches that disrupt the status quo \cite{harmon2016designing} and seek to improve the lives of foster children, such as \textit{Child-Foster Parent Matching} \cite{carnochan2013achieving,redding2000predictors,walsh1990studies}. This requires an ongoing engagement with foster parents and foster children to understand their specific values and needs as well as their cultural and parental expectations. HCI can contribute here by drawing on its rich tradition of work in action research, participatory design, and value sensitive design to incorporate the values and needs of the stakeholders \cite{fox2017social, star1999layers, asad2015illegitimate, bardzell2014utopias, borning2012next}. In addition, HCI researchers have developed methodological approaches that not only incorporate stakeholders into the design process but also the data analysis and interpretation processes \cite{baumer2017subjects, wyche2008re}. Moreover, advocating for foster children, a vulnerable and marginalized population, is inherently a social justice issue. HCI researchers have a long history of contending with social injustices and have developed theoretical and methodological approaches that seek "not so much to predict the future, but rather to imagine a radically better one \cite{fox2017social}."  Given the paucity of human-centered research into this domain and the richness of available social science literature \cite{carnochan2013achieving, redding2000predictors, blakey2012review, brown2009benefits, ryan2006investigating, cross2013children}, this presents HCI researchers with a set of complex socio-technical challenges to study.

\subsection{Recommendations for Future Research}
\vspace{0.2em}
\subsubsection{Bridging AI and HCI Through Participatory Design}
Our results indicate that there is a lack of theoretically-designed algorithms (see Table \ref{tab:characteristics}) which adds to the frustrations of child-welfare workers who are being pressured into using these algorithms as a means of standardizing decisions \cite{schwalbe2008strengthening}. This situation is further exacerbated by a lack of PD leading to algorithmic systems that offer low utility \cite{shlonsky2005next}. Only one study in our corpus engaged with child-welfare workers to understand their concerns and needs \cite{brown2019toward}. PD \cite{muller2009participatory} allows for the active inclusion of people most affected by a system. Engaging child-welfare workers in the design as well as evaluation processes ensures that their need are met and that the system integrates well with child-welfare practice. Child-welfare workers who use algorithms on a daily basis strongly stress the need to be able to explain these models to each other and to policymakers \cite{Report-LA-SDM}. Not only does this depend on which computational methods are used to construct an algorithm, how they are deployed but also how outcomes are defined and measured. This offers research pathways for HCI researchers who have increasingly started devoting attention to explaining outcomes and predictions \cite{lee2017algorithmic, miller2018see}. Moreover, it is imperative that researchers engage with the stakeholders because there are both, ethical and legal ramification of using certain types of data. For instance, legal requirements might not allow a juvenile's criminal record or history of physical and/or sexual abuse to be used for modeling \cite{shroff2017predictive}.
 
\subsubsection{Algorithmic Decision Making via Speculative Design}
We found that 56\% of studies took a deficit-based approach to mitigate risks even though child-welfare literature has discussed the significance of equitable outcomes (e.g., child-foster parent matching). Recent studies based in newer technologies still continue to focus on risk assessment and uncritically reproduce the status quo. Designing against the status quo means setting our goals beyond risk assessment, and moving more ambitiously toward design that challenges underlying problems \cite{harmon2016designing}. Human-centered speculative design \cite{baumer2017toward} can allow stakeholders to shift their focus away from algorithms and be truly innovative in how they imagine problems and their underlying causes without being constrained by what might be technologically feasible. This is especially important for algorithm design where the boundaries of possibility change every day \cite{baumer2017toward}. For instance, child-foster parent matching is well-documented in child-welfare literature for almost two decades but it has only recently been explored in an algorithm \cite{moore2016assessing} because of advances in decision-tree learning. Similarly, algorithmic advances also create novel avenues for studying the interactions and decision pathways resulting from different policies, practices and programs \cite{whittaker2017child}.

%Our results indicate that 56\% of studies took a deficit-based approach to mitigate risks with respect to bio-parents even though child-welfare literature has discussed the importance of other predictors (e.g., system factors) and outcomes (e.g., child-foster parent matching). Theoretically-founded predictors and outcomes have been inaccessible thus far because they are hard to quantify. Here, human-centered speculative design strategy \cite{baumer2017toward} can allow stakeholders to shift their focus away from algorithms and systems and be truly innovative in how they imagine problems and their supposed solutions. The fundamental idea is to not be constrained by what is technologically feasible. This is especially important for algorithm design where the boundaries of what is possible changes every day \cite{baumer2017toward}. For instance, child-foster parent matching is well-documented in child-welfare literature for almost two decades but it has only recently been developed into an algorithmic system \cite{moore2016assessing} because of advances in decision-tree learning. Policies, practices and programs have a huge impact on how children enter and remain in CWS. This creates novel avenues for studying these interactions and pathways by leveraging algorithmic advances via speculative design. 

\vspace{-0.3em}
\section{Limitations and Future Work}
We conducted a comprehensive and systematic literature review which was limited to the US-based child welfare system. We may have also missed algorithms used within CWS that are not publicly available for review. For instance, reports by non-profit organizations or state governments may have been distributed internally. Therefore, we plan to work directly with CWS agencies and conduct user interviews about the systems and algorithms being used within CWS to identify any other algorithms that have been implemented. To move towards using a human-centered approach to build new, evidence-based, and theoretically-driven algorithms, we plan to work with stakeholders in CWS to understand how different policies, practices, and programs create different decision pathways for child placements and services offered to families.

%we plan to leverage topic modeling \cite{wallach2006topic} to analyze case notes that are stored in \emph{SACWIS} to account for the contextual assessments of child-welfare workers as a salient predictor and focus our attention on outcomes that go beyond child safety concerns and improve the quality of a foster child's life. More specifically, we will focus on Child-Foster Parent matching, an outcome well-studied in social science literature that leads to child well-being.
\vspace{-0.3em}
\section{Conclusion}
In conclusion, we recommend that the HCI community partners with CWS to do the following: \textbf{1)} A renewed focus on theoretically-designed algorithms with the active engagement of stakeholders through the design and evaluation phases, \textbf{2)} Develop algorithms for practice that incorporate a more comprehensive set of predictors well-studied in child-welfare literature, as well as predictors hard to quantify thus far, and \textbf{3)} Focus on equitable outcomes founded in evidence-based child-welfare research that improve the quality of lives of foster children instead of merely mitigating future risks.

\vspace{-0.3em}
\section{Acknowledgments}
This research is funded in part by the Facebook Computational Social Science Methodology Research Award, William T. Grant Foundation (187941 and 190017), and the Northwestern Mutual Data Science Institute.

\balance{}

% REFERENCES FORMAT
% References must be the same font size as other body text.
\bibliographystyle{SIGCHI-Reference-Format}
\bibliography{sample}


\begin{thebibliography}{100}

%%% the Web address.

\ifx \showCODEN    \undefined \def \showCODEN     #1{\unskip}     \fi
\ifx \showDOI      \undefined \def \showDOI       #1{{\tt DOI:}\penalty0{#1}\ }
  \fi
\ifx \showISBNx    \undefined \def \showISBNx     #1{\unskip}     \fi
\ifx \showISBNxiii \undefined \def \showISBNxiii  #1{\unskip}     \fi
\ifx \showISSN     \undefined \def \showISSN      #1{\unskip}     \fi
\ifx \showLCCN     \undefined \def \showLCCN      #1{\unskip}     \fi
\ifx \shownote     \undefined \def \shownote      #1{#1}          \fi
\ifx \showarticletitle \undefined \def \showarticletitle #1{#1}   \fi
\ifx \showURL      \undefined \def \showURL       #1{#1}          \fi

\bibitem{HowCWSworks}
 2013.
\newblock {\em How the Child Welfare System Works}.
\newblock {T}echnical {R}eport. Children's Bureau: Child Welfare Information
  Gateway.
\newblock


\bibitem{congressbill}
 2018.
\newblock {\em Using Data To Help Protect Children and Families Act}.
\newblock 115th Congress, Senate of the United States.
\newblock


\bibitem{CRC}
 2019.
\newblock Child Welfare Goals, Legislation, and Monitoring.
\newblock   (2019).
\newblock
\showURL{%
Retrieved April 2, 2019 from
  \url{https://www.nccdglobal.org/what-we-do/children-s-research-center}}


\bibitem{aburas2018child}
{Abdurazzag~A Aburas}, {Mohammad Hassan}, {Hilary Lin}, {and} {Shreshtha
  Batshu}. 2018.
\newblock \showarticletitle{Child Maltreatment Forecast Using Bigdata
  Intelligent Approaches}. In {\em 2018 Fifth International Conference on
  Social Networks Analysis, Management and Security (SNAMS)}. IEEE, 302--308.
\newblock


\bibitem{ackerman2004organizational}
{Mark~S Ackerman} {and} {Christine Halverson}. 2004.
\newblock \showarticletitle{Organizational memory as objects, processes, and
  trajectories: An examination of organizational memory in use}.
\newblock {\em Computer Supported Cooperative Work (CSCW)\/} {13}, 2 (2004),
  155--189.
\newblock


\bibitem{alkhatib2019street}
{Ali Alkhatib} {and} {Michael Bernstein}. 2019.
\newblock \showarticletitle{Street-Level Algorithms: A Theory at the Gaps
  Between Policy and Decisions}. In {\em Proceedings of the 2019 CHI Conference
  on Human Factors in Computing Systems}. ACM, 530.
\newblock


\bibitem{andreswari2018preliminary}
{Rachmadita Andreswari}, {Irfan Darmawan}, {and} {Warih Puspitasari}. 2018.
\newblock \showarticletitle{A Preliminary Study on Detection System for
  Assessing Children and Foster Parents Suitability}. In {\em 2018 6th
  International Conference on Information and Communication Technology
  (ICoICT)}. IEEE, 376--379.
\newblock


\bibitem{asad2015illegitimate}
{Mariam Asad} {and} {Christopher~A Le~Dantec}. 2015.
\newblock \showarticletitle{Illegitimate civic participation: supporting
  community activists on the ground}. In {\em Proceedings of the 18th ACM
  Conference on Computer Supported Cooperative Work \& Social Computing}. ACM,
  1694--1703.
\newblock


\bibitem{badillo2018chibest}
{Karla Badillo-Urquiola}, {Xinru Page}, {and} {Pamela Wisniewski}. 2019.
\newblock \showarticletitle{Risk vs. Restriction: The Digital Divide between
  Providing a Sense of Normalcy and Keeping Foster Teens Safe Online}. In {\em
  Proceedings of the 2019 CHI Conference on Human Factors in Computing
  Systems}. ACM.
\newblock


\bibitem{baird1999risk}
{Christopher Baird}, {Dennis Wagner}, {Theresa Healy}, {and} {Kristen Johnson}.
  1999.
\newblock \showarticletitle{Risk assessment in child protective services:
  Consensus and actuarial model reliability}.
\newblock {\em Child Welfare\/} {78}, 6 (1999), 723.
\newblock


\bibitem{bakar2006comparative}
{Zuriana~Abu Bakar}, {Rosmayati Mohemad}, {Akbar Ahmad}, {and} {Mustafa~Mat
  Deris}. 2006.
\newblock \showarticletitle{A comparative study for outlier detection
  techniques in data mining}. In {\em 2006 IEEE conference on cybernetics and
  intelligent systems}. IEEE, 1--6.
\newblock


\bibitem{bardzell2014utopias}
{Shaowen Bardzell}. 2014.
\newblock \showarticletitle{Utopias of participation: design, criticality, and
  emancipation}. In {\em Proceedings of the 13th Participatory Design
  Conference: Short Papers, Industry Cases, Workshop Descriptions, Doctoral
  Consortium papers, and Keynote abstracts-Volume 2}. ACM, 189--190.
\newblock


\bibitem{baumer2017toward}
{Eric~PS Baumer}. 2017.
\newblock \showarticletitle{Toward human-centered algorithm design}.
\newblock {\em Big Data \& Society\/} {4}, 2 (2017), 2053951717718854.
\newblock


\bibitem{baumer2017comparing}
{Eric~PS Baumer}, {David Mimno}, {Shion Guha}, {Emily Quan}, {and} {Geri~K
  Gay}. 2017a.
\newblock \showarticletitle{Comparing grounded theory and topic modeling:
  Extreme divergence or unlikely convergence?}
\newblock {\em Journal of the Association for Information Science and
  Technology\/} {68}, 6 (2017), 1397--1410.
\newblock


\bibitem{baumer2017subjects}
{Eric~PS Baumer}, {Xiaotong Xu}, {Christine Chu}, {Shion Guha}, {and} {Geri~K
  Gay}. 2017b.
\newblock \showarticletitle{When Subjects Interpret the Data: Social Media
  Non-use as a Case for Adapting the Delphi Method to CSCW}. In {\em
  Proceedings of the 2017 ACM Conference on Computer Supported Cooperative Work
  and Social Computing}. ACM, 1527--1543.
\newblock


\bibitem{berger2009estimating}
{Lawrence~M Berger}, {Sarah~K Bruch}, {Elizabeth~I Johnson}, {Sigrid James},
  {and} {David Rubin}. 2009.
\newblock \showarticletitle{Estimating the "impact" of out-of-home placement on
  child well-being: Approaching the problem of selection bias}.
\newblock {\em Child development\/} {80}, 6 (2009), 1856--1876.
\newblock


\bibitem{blakey2012review}
{Joan~M Blakey}, {Sonya~J Leathers}, {Michelle Lawler}, {Tyreasa Washington},
  {Chiralaine Natschke}, {Tonya Strand}, {and} {Quenette Walton}. 2012.
\newblock \showarticletitle{A review of how states are addressing placement
  stability}.
\newblock {\em Children and Youth Services Review\/} {34}, 2 (2012), 369--378.
\newblock


\bibitem{borning2012next}
{Alan Borning} {and} {Michael Muller}. 2012.
\newblock \showarticletitle{Next steps for value sensitive design}. In {\em
  Proceedings of the SIGCHI conference on human factors in computing systems}.
  ACM, 1125--1134.
\newblock


\bibitem{bosk2018counts}
{Emily~Adlin Bosk}. 2018.
\newblock \showarticletitle{What counts? quantification, worker judgment, and
  divergence in child welfare decision making}.
\newblock {\em Human Service Organizations: Management, Leadership \&
  Governance\/} {42}, 2 (2018), 205--224.
\newblock


\bibitem{bozdag2013bias}
{Engin Bozdag}. 2013.
\newblock \showarticletitle{Bias in algorithmic filtering and personalization}.
\newblock {\em Ethics and information technology\/} {15}, 3 (2013), 209--227.
\newblock


\bibitem{brand1999behavior}
{Ann~E Brand} {and} {Paul~M Brinich}. 1999.
\newblock \showarticletitle{Behavior problems and mental health contacts in
  adopted, foster, and nonadopted children}.
\newblock {\em The journal of child psychology and psychiatry and allied
  disciplines\/} {40}, 8 (1999), 1221--1229.
\newblock


\bibitem{braun2006using}
{Virginia Braun} {and} {Victoria Clarke}. 2006.
\newblock \showarticletitle{Using thematic analysis in psychology}.
\newblock {\em Qualitative research in psychology\/} {3}, 2 (2006), 77--101.
\newblock


\bibitem{breiman2001statistical}
{Leo Breiman} {and} {others}. 2001.
\newblock \showarticletitle{Statistical modeling: The two cultures (with
  comments and a rejoinder by the author)}.
\newblock {\em Statistical science\/} {16}, 3 (2001), 199--231.
\newblock


\bibitem{brindley2018can}
{Meredith Brindley}, {James~P Heyes}, {and} {Darrell Booker}. 2018.
\newblock \showarticletitle{Can Machine Learning Create an Advocate for Foster
  Youth?}
\newblock {\em Journal of Technology in Human Services\/} {36}, 1 (2018),
  31--36.
\newblock


\bibitem{brown2019toward}
{Anna Brown}, {Alexandra Chouldechova}, {Emily Putnam-Hornstein}, {Andrew
  Tobin}, {and} {Rhema Vaithianathan}. 2019.
\newblock \showarticletitle{Toward Algorithmic Accountability in Public
  Services: A Qualitative Study of Affected Community Perspectives on
  Algorithmic Decision-making in Child Welfare Services}. In {\em Proceedings
  of the 2019 CHI Conference on Human Factors in Computing Systems}. ACM, 41.
\newblock


\bibitem{brown2009benefits}
{Jason~D Brown}, {Natalie George}, {Jennifer Sintzel}, {and} {David~St
  Arnault}. 2009.
\newblock \showarticletitle{Benefits of cultural matching in foster care}.
\newblock {\em Children and Youth Services Review\/} {31}, 9 (2009),
  1019--1024.
\newblock


\bibitem{camasso1995prediction}
{Michael~J Camasso} {and} {Radha Jagannathan}. 1995.
\newblock \showarticletitle{Prediction accuracy of the Washington and Illinois
  risk assessment instruments: An application of receiver operating
  characteristic curve analysis}.
\newblock {\em Social Work Research\/} {19}, 3 (1995), 174--183.
\newblock


\bibitem{camasso2000modeling}
{Michael~J Camasso} {and} {Radha Jagannathan}. 2000.
\newblock \showarticletitle{Modeling the reliability and predictive validity of
  risk assessment in child protective services}.
\newblock {\em Children and Youth Services Review\/} {22}, 11-12 (2000),
  873--896.
\newblock


\bibitem{camasso2013decision}
{Michael~J Camasso} {and} {Radha Jagannathan}. 2013.
\newblock \showarticletitle{Decision making in child protective services: A
  risky business?}
\newblock {\em Risk analysis\/} {33}, 9 (2013), 1636--1649.
\newblock


\bibitem{carnochan2013achieving}
{Sarah Carnochan}, {Megan Moore}, {and} {Michael~J Austin}. 2013.
\newblock \showarticletitle{Achieving placement stability}.
\newblock {\em Journal of Evidence-Based Social Work\/} {10}, 3 (2013),
  235--253.
\newblock


\bibitem{chancellor2016quantifying}
{Stevie Chancellor}, {Zhiyuan Lin}, {Erica~L Goodman}, {Stephanie Zerwas},
  {and} {Munmun De~Choudhury}. 2016.
\newblock \showarticletitle{Quantifying and predicting mental illness severity
  in online pro-eating disorder communities}. In {\em Proceedings of the 19th
  ACM Conference on Computer-Supported Cooperative Work \& Social Computing}.
  ACM, 1171--1184.
\newblock


\bibitem{chor2012predicting}
{Ka~Ho~Brian Chor}, {Gary~M McClelland}, {Dana~A Weiner}, {Neil Jordan}, {and}
  {John~S Lyons}. 2012.
\newblock \showarticletitle{Predicting outcomes of children in residential
  treatment: A comparison of a decision support algorithm and a
  multidisciplinary team decision model}.
\newblock {\em Children and Youth Services Review\/} {34}, 12 (2012),
  2345--2352.
\newblock


\bibitem{chouldechova2018case}
{Alexandra Chouldechova}, {Diana Benavides-Prado}, {Oleksandr Fialko}, {and}
  {Rhema Vaithianathan}. 2018.
\newblock \showarticletitle{A case study of algorithm-assisted decision making
  in child maltreatment hotline screening decisions}. In {\em Conference on
  Fairness, Accountability and Transparency}. 134--148.
\newblock


\bibitem{church2017search}
{Christopher~E Church} {and} {Amanda~J Fairchild}. 2017.
\newblock \showarticletitle{In Search of a Silver Bullet: Child Welfare's
  Embrace of Predictive Analytics}.
\newblock {\em Juvenile and Family Court Journal\/} {68}, 1 (2017), 67--81.
\newblock


\bibitem{clifford1986writing}
{James Clifford} {and} {George~E Marcus}. 1986.
\newblock {\em Writing culture: The poetics and politics of ethnography}.
\newblock Univ of California Press.
\newblock


\bibitem{cohen2014applied}
{Patricia Cohen}, {Stephen~G West}, {and} {Leona~S Aiken}. 2014.
\newblock {\em Applied multiple regression/correlation analysis for the
  behavioral sciences}.
\newblock Psychology Press.
\newblock


\bibitem{congress2008fostering}
{US Congress}. 2008.
\newblock Fostering connections to success and increasing adoptions act of
  2008.
\newblock   (2008).
\newblock


\bibitem{cooper2005implications}
{Lindsay~D Cooper}. 2005.
\newblock \showarticletitle{Implications of media scrutiny for a child
  protection agency}.
\newblock {\em J. Soc. \& Soc. Welfare\/}  {32} (2005), 107.
\newblock


\bibitem{cordell2016patterns}
{Katharan~D Cordell}, {Lonnie~R Snowden}, {and} {Laura Hosier}. 2016.
\newblock \showarticletitle{Patterns and priorities of service need identified
  through the Child and Adolescent Needs and Strengths (CANS) assessment}.
\newblock {\em Children and Youth Services Review\/}  {60} (2016), 129--135.
\newblock


\bibitem{corrigan_2019}
{Michael Corrigan}. 2019.
\newblock Building A Comprehensive Child Welfare Information System.
\newblock   (Jan 2019).
\newblock
\showURL{%
\url{https://chronicleofsocialchange.org/child-welfare-2/building-comprehensive-child-welfare-information-system/33426}}


\bibitem{cross2013children}
{Theodore~P Cross}, {EUN Koh}, {Nancy Rolock}, {and} {Jennifer Eblen-Manning}.
  2013.
\newblock \showarticletitle{Why do children experience multiple placement
  changes in foster care? Content analysis on reasons for instability}.
\newblock {\em Journal of Public Child Welfare\/} {7}, 1 (2013), 39--58.
\newblock


\bibitem{d2008risk}
{Amy D'andrade}, {Michael~J Austin}, {and} {Amy Benton}. 2008.
\newblock \showarticletitle{Risk and safety assessment in child welfare:
  Instrument comparisons}.
\newblock {\em Journal of evidence-based social work\/} {5}, 1-2 (2008),
  31--56.
\newblock


\bibitem{danks2017algorithmic}
{David Danks} {and} {Alex~John London}. 2017.
\newblock \showarticletitle{Algorithmic Bias in Autonomous Systems.}. In {\em
  IJCAI}. 4691--4697.
\newblock


\bibitem{danktert2013risk}
{EW Danktert} {and} {Kristen Johnson}. 2013.
\newblock \showarticletitle{Risk assessment validation: A prospective study}.
\newblock {\em Los Angeles: California Department of Social Services, Children
  and Family Services Division\/} (2013).
\newblock


\bibitem{davoren1975foster}
{Elizabeth Davoren}. 1975.
\newblock \showarticletitle{Foster placement of abused children}.
\newblock {\em Children today\/} {4}, 3 (1975), 41.
\newblock


\bibitem{de2018integrating}
{Munmun De~Choudhury} {and} {Emre Kiciman}. 2018.
\newblock \showarticletitle{Integrating Artificial and Human Intelligence in
  Complex, Sensitive Problem Domains: Experiences from Mental Health}.
\newblock {\em AI Magazine\/} {39}, 3 (2018), 69--80.
\newblock


\bibitem{dettlaff2011disentangling}
{Alan~J Dettlaff}, {Stephanie~L Rivaux}, {Donald~J Baumann}, {John~D Fluke},
  {Joan~R Rycraft}, {and} {Joyce James}. 2011.
\newblock \showarticletitle{Disentangling substantiation: The influence of
  race, income, and risk on the substantiation decision in child welfare}.
\newblock {\em Children and Youth Services Review\/} {33}, 9 (2011),
  1630--1637.
\newblock


\bibitem{donoho201750}
{David Donoho}. 2017.
\newblock \showarticletitle{50 years of data science}.
\newblock {\em Journal of Computational and Graphical Statistics\/} {26}, 4
  (2017), 745--766.
\newblock


\bibitem{fluke2010placement}
{John~D Fluke}, {Martin Chabot}, {Barbara Fallon}, {Bruce MacLaurin}, {and}
  {Cindy Blackstock}. 2010.
\newblock \showarticletitle{Placement decisions and disparities among
  aboriginal groups: An application of the decision making ecology through
  multi-level analysis}.
\newblock {\em Child Abuse \& Neglect\/} {34}, 1 (2010), 57--69.
\newblock


\bibitem{acf_ccwis}
{Administration for Children} {and} {Families}.
\newblock {\em Comprehensive Child Welfare Information System\/} (106 ed.).
  Vol.~81.
\newblock Federal Register: The Daily Journal of the United States.
\newblock


\bibitem{fowler2019scaling}
{Patrick~J Fowler}, {Katherine~E Marcal}, {Saras Chung}, {Derek~S Brown},
  {Melissa Jonson-Reid}, {and} {Peter~S Hovmand}. 2019.
\newblock \showarticletitle{Scaling up housing services within the child
  welfare system: policy insights from simulation modeling}.
\newblock {\em Child maltreatment\/} (2019), 1077559519846431.
\newblock


\bibitem{fox2017social}
{Sarah Fox}, {Jill Dimond}, {Lilly Irani}, {Tad Hirsch}, {Michael Muller},
  {and} {Shaowen Bardzell}. 2017.
\newblock \showarticletitle{Social Justice and Design: Power and oppression in
  collaborative systems}. In {\em Companion of the 2017 ACM Conference on
  Computer Supported Cooperative Work and Social Computing}. ACM, 117--122.
\newblock


\bibitem{furr2017psychometrics}
{R~Michael Furr}. 2017.
\newblock {\em Psychometrics: an introduction}.
\newblock Sage Publications.
\newblock


\bibitem{gambrill2000risk}
{Eileen Gambrill} {and} {Aron Shlonsky}. 2000.
\newblock Risk assessment in context.
\newblock   (2000).
\newblock


\bibitem{gambrill2001need}
{Eileen Gambrill} {and} {Aron Shlonsky}. 2001.
\newblock \showarticletitle{The need for comprehensive risk management systems
  in child welfare}.
\newblock {\em Children and Youth Services Review\/} {23}, 1 (2001), 79--107.
\newblock


\bibitem{gray2019trove}
{Stuart Gray}, {Kirsten Cater}, {Chloe Meineck}, {Rachel Hahn}, {Debbie
  Watson}, {and} {Tom Metcalfe}. 2019.
\newblock \showarticletitle{trove: A digitally enhanced memory box for looked
  after and adopted children}. In {\em Proceedings of the 18th ACM
  International Conference on Interaction Design and Children}. ACM, 458--463.
\newblock


\bibitem{gregory2012structured}
{Robin Gregory}, {Lee Failing}, {Michael Harstone}, {Graham Long}, {Tim
  McDaniels}, {and} {Dan Ohlson}. 2012.
\newblock {\em Structured decision making: a practical guide to environmental
  management choices}.
\newblock John Wiley \& Sons.
\newblock


\bibitem{harmon2016designing}
{Ellie Harmon}, {Matthias Korn}, {Ann Light}, {and} {Amy Voida}. 2016.
\newblock \showarticletitle{Designing against the status quo}. In {\em
  Proceedings of the 2016 ACM Conference Companion Publication on Designing
  Interactive Systems}. ACM, 65--68.
\newblock


\bibitem{harrison2018tale}
{Teresa~M Harrison}, {Donna Canestraro}, {Theresa Pardo}, {Martha
  Avila-Marilla}, {Nicolas Soto}, {Megan Sutherland}, {Brian Burke}, {and}
  {Mila Gasco}. 2018.
\newblock \showarticletitle{A tale of two information systems: transitioning to
  a data-centric information system for child welfare}. In {\em Proceedings of
  the 19th Annual International Conference on Digital Government Research:
  Governance in the Data Age}. ACM, 108.
\newblock


\bibitem{inkpen2019human}
{Kori Inkpen}, {Stevie Chancellor}, {Munmun De~Choudhury}, {Michael Veale},
  {and} {Eric~PS Baumer}. 2019.
\newblock \showarticletitle{Where is the Human?: Bridging the Gap Between AI
  and HCI}. In {\em Extended Abstracts of the 2019 CHI Conference on Human
  Factors in Computing Systems}. ACM, W09.
\newblock


\bibitem{luma2012innovating}
{Luma Institute}. 2012.
\newblock {\em Innovating for people: Handbook of human-centered design
  methods}.
\newblock LUMA Institute, LLC.
\newblock


\bibitem{dcf_illinois}
{David Jackson} {and} {Gary Marx}. 2017.
\newblock Data mining program designed to predict child abuse proves
  unreliable, DCFS says.
\newblock   (dec 2017).
\newblock
\showURL{%
\url{https://www.chicagotribune.com/investigations/ct-dcfs-eckerd-met-20171206-story.html}}


\bibitem{james2004foster}
{Sigrid James}. 2004.
\newblock \showarticletitle{Why do foster care placements disrupt? An
  investigation of reasons for placement change in foster care}.
\newblock {\em Social service review\/} {78}, 4 (2004), 601--627.
\newblock


\bibitem{johnson2003california}
{K Johnson} {and} {D Wagner}. 2003.
\newblock California Structured Decision Making. Risk Assessment Revalidation:
  A Prospective Study. Children’s Research Center. SAMSHA’s National
  Registry of Evidence-based Programs and Practices (NREPP).
\newblock   (2003).
\newblock


\bibitem{johnson2004effectiveness}
{Will Johnson}. 2004.
\newblock \showarticletitle{Effectiveness of California’s child welfare
  structured decision making (SDM) model: a prospective study of the validity
  of the California Family Risk Assessment}.
\newblock {\em Madison (Wisconsin, USA): Children’s Research Center\/}
  (2004).
\newblock


\bibitem{kleinberg2015prediction}
{Jon Kleinberg}, {Jens Ludwig}, {Sendhil Mullainathan}, {and} {Ziad Obermeyer}.
  2015.
\newblock \showarticletitle{Prediction policy problems}.
\newblock {\em American Economic Review\/} {105}, 5 (2015), 491--95.
\newblock


\bibitem{koh2008propensity}
{Eun Koh} {and} {Mark~F Testa}. 2008.
\newblock \showarticletitle{Propensity score matching of children in kinship
  and nonkinship foster care: Do permanency outcomes differ?}
\newblock {\em Social Work Research\/} {32}, 2 (2008), 105--116.
\newblock


\bibitem{kutz2013data}
{J~Nathan Kutz}. 2013.
\newblock {\em Data-driven modeling \& scientific computation: methods for
  complex systems \& big data}.
\newblock Oxford University Press.
\newblock


\bibitem{lambrecht2019algorithmic}
{Anja Lambrecht} {and} {Catherine Tucker}. 2019.
\newblock \showarticletitle{Algorithmic Bias? An Empirical Study of Apparent
  Gender-Based Discrimination in the Display of STEM Career Ads}.
\newblock {\em Management Science\/} (2019).
\newblock


\bibitem{lardner2015restrictiveness}
{Mark~D Lardner}. 2015.
\newblock \showarticletitle{Are restrictiveness of care decisions based on
  youth level of need? A multilevel model analysis of placement levels using
  the child and adolescent needs and strengths assessment}.
\newblock {\em Residential Treatment for Children \& Youth\/} {32}, 3 (2015),
  195--207.
\newblock


\bibitem{lee2017algorithmic}
{Min~Kyung Lee} {and} {Su Baykal}. 2017.
\newblock \showarticletitle{Algorithmic mediation in group decisions: Fairness
  perceptions of algorithmically mediated vs. discussion-based social
  division}. In {\em Proceedings of the 2017 ACM Conference on Computer
  Supported Cooperative Work and Social Computing}. ACM, 1035--1048.
\newblock


\bibitem{lodato2018institutional}
{Thomas Lodato} {and} {Carl DiSalvo}. 2018.
\newblock \showarticletitle{Institutional constraints: the forms and limits of
  participatory design in the public realm}. In {\em Proceedings of the 15th
  Participatory Design Conference: Full Papers-Volume 1}. ACM, 5.
\newblock


\bibitem{lyons2004measurement}
{John~S Lyons}, {Dana~Aron Weiner}, {and} {Melanie~Buddin Lyons}. 2004.
\newblock \showarticletitle{Measurement as communication in outcomes
  management: The child and adolescent needs and strengths (CANS)}.
\newblock {\em The Use of Psychological Testing for Treatment Planning and
  Outcomes Assessment. Volume 2: Instruments for Children and Adolescents\/}
  (2004).
\newblock


\bibitem{mackenzie2011toward}
{Michael~J MacKenzie}, {Jonathan~B Kotch}, {and} {Li-Ching Lee}. 2011.
\newblock \showarticletitle{Toward a cumulative ecological risk model for the
  etiology of child maltreatment}.
\newblock {\em Children and youth services review\/} {33}, 9 (2011),
  1638--1647.
\newblock


\bibitem{maclean2015forum77}
{Diana MacLean}, {Sonal Gupta}, {Anna Lembke}, {Christopher Manning}, {and}
  {Jeffrey Heer}. 2015.
\newblock \showarticletitle{Forum77: An analysis of an online health forum
  dedicated to addiction recovery}. In {\em Proceedings of the 18th ACM
  Conference on Computer Supported Cooperative Work \& Social Computing}. ACM,
  1511--1526.
\newblock


\bibitem{march1991exploration}
{James~G March}. 1991.
\newblock \showarticletitle{Exploration and exploitation in organizational
  learning}.
\newblock {\em Organization science\/} {2}, 1 (1991), 71--87.
\newblock


\bibitem{marshall2000neural}
{David~B Marshall} {and} {Diana~J English}. 2000.
\newblock \showarticletitle{Neural network modeling of risk assessment in child
  protective services.}
\newblock {\em Psychological Methods\/} {5}, 1 (2000), 102.
\newblock


\bibitem{mcdonald2002predicting}
{Thomas~P McDonald}, {John Poertner}, {and} {Gardenia Harris}. 2002.
\newblock \showarticletitle{Predicting placement in foster care: A comparison
  of logistic regression and neural network analysis}.
\newblock {\em Journal of social service research\/} {28}, 2 (2002), 1--20.
\newblock


\bibitem{miller2018see}
{Hannah Miller~Hillberg}, {Zachary Levonian}, {Daniel Kluver}, {Loren Terveen},
  {and} {Brent Hecht}. 2018.
\newblock \showarticletitle{What I See is What You Don't Get: The Effects of
  (Not) Seeing Emoji Rendering Differences across Platforms}.
\newblock {\em Proceedings of the ACM on Human-Computer Interaction\/} {2},
  CSCW (2018), 124.
\newblock


\bibitem{moore2016assessing}
{Terry~D Moore}, {Thomas~P McDonald}, {and} {Kari Cronbaugh-Auld}. 2016.
\newblock \showarticletitle{Assessing risk of placement instability to aid
  foster care placement decision making}.
\newblock {\em Journal of Public Child Welfare\/} {10}, 2 (2016), 117--131.
\newblock


\bibitem{muller2016machine}
{Michael Muller}, {Shion Guha}, {Eric~PS Baumer}, {David Mimno}, {and} {N~Sadat
  Shami}. 2016.
\newblock \showarticletitle{Machine learning and grounded theory method:
  Convergence, divergence, and combination}. In {\em Proceedings of the 19th
  International Conference on Supporting Group Work}. ACM, 3--8.
\newblock


\bibitem{muller2009participatory}
{Michael~J Muller}. 2009.
\newblock \showarticletitle{Participatory design: the third space in HCI}.
\newblock In {\em Human-computer interaction}. CRC press, 181--202.
\newblock


\bibitem{Report-LA-SDM}
{Judge~Michael Nash}. 2017.
\newblock \showarticletitle{Examination of using Structured Decision Making and
  Predictive Analytics in assessing Safety and Risk in Child Welfare}.
\newblock {\em County of Los Angeles Office of Child Protection\/} (May 2017).
\newblock


\bibitem{needell2003black}
{Barbara Needell}, {M~Alan Brookhart}, {and} {Seon Lee}. 2003.
\newblock \showarticletitle{Black children and foster care placement in
  California}.
\newblock {\em Children and Youth Services Review\/} {25}, 5-6 (2003),
  393--408.
\newblock


\bibitem{noonan2009legal}
{Kathleen~G Noonan}, {Charles~F Sabel}, {and} {William~H Simon}. 2009.
\newblock \showarticletitle{Legal accountability in the service-based welfare
  state: Lessons from child welfare reform}.
\newblock {\em Law \& Social Inquiry\/} {34}, 3 (2009), 523--568.
\newblock


\bibitem{acf2017}
{US~Department of Health} {and} {Human Services}. 2017.
\newblock \showarticletitle{Child Maltreatment 2017}.
\newblock {\em Children's Bureau (Ed.)\/} (2017).
\newblock


\bibitem{us2017afcars}
{US~Department of Health}, {Human Services}, {and} {others}. 2017.
\newblock \showarticletitle{The AFCARS report: Preliminary FY 2016 estimates as
  of Oct 2017}.
\newblock {\em Children's Bureau (Ed.)\/}  {21} (2017), 6.
\newblock


\bibitem{ott2015introduction}
{R~Lyman Ott} {and} {Micheal~T Longnecker}. 2015.
\newblock {\em An introduction to statistical methods and data analysis}.
\newblock Nelson Education.
\newblock


\bibitem{pink2013applying}
{Sarah Pink}, {Kerstin~Leder Mackley}, {Val Mitchell}, {Marcus Hanratty},
  {Carolina Escobar-Tello}, {Tracy Bhamra}, {and} {Roxana Morosanu}. 2013.
\newblock \showarticletitle{Applying the lens of sensory ethnography to
  sustainable HCI}.
\newblock {\em ACM Transactions on Computer-Human Interaction (TOCHI)\/} {20},
  4 (2013), 25.
\newblock


\bibitem{pinter2017adolescent}
{Anthony~T Pinter}, {Pamela~J Wisniewski}, {Heng Xu}, {Mary~Beth Rosson}, {and}
  {Jack~M Caroll}. 2017.
\newblock \showarticletitle{Adolescent online safety: Moving beyond formative
  evaluations to designing solutions for the future}. In {\em Proceedings of
  the 2017 Conference on Interaction Design and Children}. ACM, 352--357.
\newblock


\bibitem{putnam2013racial}
{Emily Putnam-Hornstein}, {Barbara Needell}, {Bryn King}, {and} {Michelle
  Johnson-Motoyama}. 2013.
\newblock \showarticletitle{Racial and ethnic disparities: A population-based
  examination of risk factors for involvement with child protective services}.
\newblock {\em Child Abuse \& Neglect\/} {37}, 1 (2013), 33--46.
\newblock


\bibitem{redding2000predictors}
{Richard~E Redding}, {Carrie Fried}, {and} {Preston~A Britner}. 2000.
\newblock \showarticletitle{Predictors of placement outcomes in treatment
  foster care: Implications for foster parent selection and service delivery}.
\newblock {\em Journal of child and family studies\/} {9}, 4 (2000), 425--447.
\newblock


\bibitem{ringel2018improving}
{Jeanne~S Ringel}, {Dana Schultz}, {Joshua Mendelsohn}, {Stephanie~Brooks
  Holliday}, {Katharine Sieck}, {Ifeanyi Edochie}, {and} {Lauren Davis}. 2018.
\newblock \showarticletitle{Improving child welfare outcomes: balancing
  investments in prevention and treatment}.
\newblock {\em Rand health quarterly\/} {7}, 4 (2018).
\newblock


\bibitem{ryan2006investigating}
{Joseph~P Ryan}, {Philip Garnier}, {Michael Zyphur}, {and} {Fuhua Zhai}. 2006.
\newblock \showarticletitle{Investigating the effects of caseworker
  characteristics in child welfare}.
\newblock {\em Children and Youth Services Review\/} {28}, 9 (2006), 993--1006.
\newblock


\bibitem{schuerman1986computer}
{John~R Schuerman} {and} {Lynn~Harold Vogel}. 1986.
\newblock \showarticletitle{Computer support of placement planning: the use of
  expert systems in child welfare.}
\newblock {\em Child welfare\/} {65}, 6 (1986), 531--543.
\newblock


\bibitem{schwab1989continuum}
{A~James Schwab} {and} {Susan~S Wilson}. 1989.
\newblock \showarticletitle{The continuum of care system: Decision support for
  practitioners}.
\newblock {\em Computers in Human Services\/} {4}, 1-2 (1989), 123--140.
\newblock


\bibitem{schwab1984matching}
{A~James Schwab~Jr}, {Michael~E Bruce}, {and} {Ruth~G McRoy}. 1984.
\newblock \showarticletitle{Matching children with placements}.
\newblock {\em Children and youth services review\/} {6}, 2 (1984), 125--133.
\newblock


\bibitem{schwalbe2004re}
{Craig Schwalbe}. 2004.
\newblock \showarticletitle{Re-visioning risk assessment for human service
  decision making}.
\newblock {\em Children and Youth Services Review\/} {26}, 6 (2004), 561--576.
\newblock


\bibitem{schwalbe2008strengthening}
{Craig~S Schwalbe}. 2008.
\newblock \showarticletitle{Strengthening the integration of actuarial risk
  assessment with clinical judgment in an evidence based practice framework}.
\newblock {\em Children and Youth Services Review\/} {30}, 12 (2008),
  1458--1464.
\newblock


\bibitem{shlonsky2005next}
{Aron Shlonsky} {and} {Dennis Wagner}. 2005.
\newblock \showarticletitle{The next step: Integrating actuarial risk
  assessment and clinical judgment into an evidence-based practice framework in
  CPS case management}.
\newblock {\em Children and youth services review\/} {27}, 4 (2005), 409--427.
\newblock


\bibitem{shroff2017predictive}
{Ravi Shroff}. 2017.
\newblock \showarticletitle{Predictive Analytics for City Agencies: Lessons
  from Children's Services}.
\newblock {\em Big data\/} {5}, 3 (2017), 189--196.
\newblock


\bibitem{sicoly1989computer}
{Fiore Sicoly}. 1989a.
\newblock \showarticletitle{Computer-aided decisions in human services: Expert
  systems and multivariate models}.
\newblock {\em Computers in Human Behavior\/} {5}, 1 (1989), 47--60.
\newblock


\bibitem{sicoly1989prediction}
{Fiore Sicoly}. 1989b.
\newblock \showarticletitle{Prediction and decision making in child welfare}.
\newblock {\em Computers in Human Services\/} {5}, 3-4 (1989), 43--56.
\newblock


\bibitem{simpson2000statistical}
{Douglas~G Simpson}, {Peter~B Imrey}, {Olga Geling}, {and} {Susan Butkus}.
  2000.
\newblock \showarticletitle{Statistical estimation of child abuse rates from
  administrative databases}.
\newblock {\em Children and Youth Services Review\/} {22}, 11-12 (2000),
  951--971.
\newblock


\bibitem{star1999layers}
{Susan~Leigh Star} {and} {Anselm Strauss}. 1999.
\newblock \showarticletitle{Layers of silence, arenas of voice: The ecology of
  visible and invisible work}.
\newblock {\em Computer supported cooperative work (CSCW)\/} {8}, 1-2 (1999),
  9--30.
\newblock


\bibitem{stevens1984outliers}
{James~P Stevens}. 1984.
\newblock \showarticletitle{Outliers and influential data points in regression
  analysis.}
\newblock {\em Psychological Bulletin\/} {95}, 2 (1984), 334.
\newblock


\bibitem{stone1983prediction}
{Norman~M Stone} {and} {Susan~F Stone}. 1983.
\newblock \showarticletitle{The prediction of successful foster placement}.
\newblock {\em Social Casework\/} {64}, 1 (1983), 11--17.
\newblock


\bibitem{strohmayer2015exploring}
{Angelika Strohmayer}, {Rob Comber}, {and} {Madeline Balaam}. 2015.
\newblock \showarticletitle{Exploring learning ecologies among people
  experiencing homelessness}. In {\em Proceedings of the 33rd Annual ACM
  Conference on Human Factors in Computing Systems}. ACM, 2275--2284.
\newblock


\bibitem{tushnet2018difference}
{Rebecca Tushnet}. 2018.
\newblock \showarticletitle{The Difference Engine: Perpetuating Poverty through
  Algorithms}.
\newblock {\em Jotwell: J. Things We Like\/} (2018), 1.
\newblock


\bibitem{vaithianathan2017developing}
{Rhema Vaithianathan}, {Emily Putnam-Hornstein}, {Nan Jiang}, {Parma Nand},
  {and} {Tim Maloney}. 2017.
\newblock \showarticletitle{Developing predictive models to support child
  maltreatment hotline screening decisions: Allegheny County methodology and
  implementation}.
\newblock {\em Center for Social data Analytics\/} (2017).
\newblock


\bibitem{wagner1998using}
{D Wagner}, {K Johnson}, {and} {W Johnson}. 1998.
\newblock \showarticletitle{Using actuarial risk assessment to target service
  nterventions in pilot California Counties}.
\newblock {\em 13th National Roundtable on CPS Risk Assessment, San Francisco,
  CA\/} (1998).
\newblock


\bibitem{wallach2006topic}
{Hanna~M Wallach}. 2006.
\newblock \showarticletitle{Topic modeling: beyond bag-of-words}. In {\em
  Proceedings of the 23rd international conference on Machine learning}. ACM,
  977--984.
\newblock


\bibitem{walsh1990studies}
{James~A Walsh} {and} {Roberta~A Walsh}. 1990.
\newblock \showarticletitle{Studies of the maintenance of subsidized foster
  placements in the Casey Family Program.}
\newblock {\em Child Welfare\/} {69}, 2 (1990), 99--114.
\newblock


\bibitem{webster2000placement}
{Daniel Webster}, {Richard~P Barth}, {and} {Barbara Needell}. 2000.
\newblock \showarticletitle{Placement stability for children in out-of-home
  care: A longitudinal analysis}.
\newblock {\em CHILD WELFARE-NEW YORK-\/} {79}, 5 (2000), 614--632.
\newblock


\bibitem{webster2002analyzing}
{Jane Webster} {and} {Richard~T Watson}. 2002.
\newblock \showarticletitle{Analyzing the past to prepare for the future:
  Writing a literature review}.
\newblock {\em MIS quarterly\/} (2002), xiii--xxiii.
\newblock


\bibitem{whittaker2017child}
{James~K Whittaker}. 2017.
\newblock {\em The child welfare challenge: Policy, practice, and research}.
\newblock Routledge.
\newblock


\bibitem{wobbrock2016research}
{Jacob~O Wobbrock} {and} {Julie~A Kientz}. 2016.
\newblock \showarticletitle{Research contributions in human-computer
  interaction}.
\newblock {\em interactions\/} {23}, 3 (2016), 38--44.
\newblock


\bibitem{woelfer2010homeless}
{Jill~Palzkill Woelfer} {and} {David~G Hendry}. 2010.
\newblock \showarticletitle{Homeless young people's experiences with
  information systems: life and work in a community technology center}. In {\em
  Proceedings of the SIGCHI Conference on Human Factors in Computing Systems}.
  ACM, 1291--1300.
\newblock


\bibitem{wyche2008re}
{Susan~P Wyche}, {Paul~M Aoki}, {and} {Rebecca~E Grinter}. 2008.
\newblock \showarticletitle{Re-placing faith: reconsidering the
  secular-religious use divide in the United States and Kenya}. In {\em
  Proceedings of the SIGCHI conference on human factors in computing systems}.
  ACM, 11--20.
\newblock


\bibitem{zuravin1999child}
{Susan~J Zuravin}. 1999.
\newblock \showarticletitle{Child neglect}.
\newblock {\em Neglected children: Research, practice, and policy\/} (1999),
  24--46.
\newblock


\end{thebibliography}

\end{document}